\documentclass[12pt]{article}
\usepackage{jheppub}
\usepackage{amsmath,amssymb,amsfonts,graphicx,slashed,amsthm,mathtools,upgreek, enumerate, tensor, subfig,tikz}
\usepackage[dvipsnames]{xcolor}
\usepackage{arydshln}

\usepackage{comment}
\usepackage{hyperref}
\usepackage[utf8]{inputenc}
\usepackage[titletoc]{appendix}
\usepackage{esvect}

\numberwithin{equation}{section} 
\allowdisplaybreaks

\allowdisplaybreaks

\newcommand{\be}{\begin{equation}}
\newcommand{\ee}{\end{equation}}
\newcommand{\f}{\frac}

\newcommand{\bea}{\begin{eqnarray}}
\newcommand{\eea}{\end{eqnarray}}
\newcommand{\ba}{\begin{align}}
\newcommand{\ea}{\end{align}}

\newcommand{\la}{\langle}
\newcommand{\ra}{\rangle}
\newcommand{\beq}{\begin{equation}}
\newcommand{\eeq}{\end{equation}}
\newcommand{\bra}[1]{\langle #1 |}
\newcommand{\ket}[1]{| #1 \rangle}

\DeclareMathOperator{\tr}{tr}

\newcommand{\hm}{\mathcal{H}}

\newcommand{\ml}{\mathcal{L}}
\newcommand{\vN}{\overrightarrow{\mathcal N}}
\newcommand{\al}{\alpha}

\title{
Emergent Closed Universes in Symmetric Orbifold CFTs
}

\author[a]{Akihiro Miyata}
\author[b]{\!Keigo Horikoshi}
\author[c]{\!Hiromasa Tajima}
\author[b]{\! Tomonori Ugajin}

\affiliation[\,a]{Center for Gravitational Physics and Quantum Information (CGPQI), \\
Yukawa Institute for Theoretical Physics (YITP), Kyoto University,\\
	Kitashirakawa Oiwakecho, 
    Sakyo-ku, Kyoto 606-8502, Japan}
\affiliation[\,b]{Department of Physics, Rikkyo University, Toshima, Tokyo 171-8501, Japan}
\affiliation[\,c]{Department of Physics, Graduate School of Science, Nagoya University, Nagoya 464-8602, Japan}

\emailAdd{akihiro.miyata@yukawa.kyoto-u.ac.jp}
\emailAdd{24ra002c@rikkyo.ac.jp}
\emailAdd{tajima.hiromasa.m6@s.mail.nagoya-u.ac.jp}
\emailAdd{ugajin@rikkyo.ac.jp}

\abstract{We identify closed universe sectors in large $N$ symmetric orbifold CFTs with holographic duals. Starting from tensor product states built out of a finite dimensional low energy subspace of the seed theory, we show that the large $N$ Hilbert space decomposes into superselection sectors labeled by occupation number distributions. Before imposing the orbifold gauge constraint, these sectors have exponentially large dimensions, and the maximally entropic sector dominates the ungauged Hilbert space. We argue that this sector exhibits several characteristic features expected of a closed universe Hilbert space: pure states become indistinguishable from a mixed state at the level of simple correlation functions, and the associated operator algebra  is naturally a hyperfinite type II$_1$ von Neumann algebra. We then impose the $S_N$ gauge constraint. The large gauge redundancy drastically reduces the number of independent states. In particular, in the large $N$ limit,  the dimension of the physical Hilbert space grows only polynomially with $N$. Consequently, each superselection sector after imposing the constraint is one dimensional in this limit. This reproduces the qualitative behavior suggested by gravitational path integral calculations with wormholes.
We then show why, in this setup, the Hartle-Hawking type semiclassical approximation for the dominant closed universe fails to reproduce the CFT results. Nevertheless, the dominant saddle point approximation for gravitational path integral calculation is reconstructed once the CFT degrees of freedom are coupled to external observer degrees of freedom.

} 

\keywords{}

\begin{document}

\maketitle

\section{Introduction}

Closed universes raise a number of conceptual puzzles in quantum gravity \cite{Almheiri:2020cfm,Chen:2020tes,Hartman:2020khs,Balasubramanian:2020xqf}. One particularly sharp issue concerns the dimensions of their Hilbert spaces. The calculation of the Gram matrix by gravitational path integral calculations including wormhole contributions reveals that the rank of this matrix is one, leading to one dimensional Hilbert space~\cite{Hsin:2020mfa}. Motivated by this puzzle, recent discussions have emphasized the importance of introducing observer degrees of freedom in order to define nontrivial physics in a closed universe~\cite{PhysRevD.27.2885,Chandrasekaran:2022cip,Balasubramanian:2023xyd,Usatyuk:2024isz,
Harlow:2025pvj}. Another closely related, though conceptually distinct, formulation of semiclassical physics is based on conditioning on a future time slice ~\cite{Nomura:2025whc,Nomura:2026igt}.

At present, AdS/CFT correspondence \cite{Maldacena:1997re} provides our best understood nonperturbative framework for quantum gravity, and therefore offers a controlled setting in which one can ask how a closed universe is encoded in a microscopic CFT description. 

Important progress in this direction has been made in recent works.
In \cite{Antonini:2023hdh}, asymptotically two sided AdS solution containing an additional disjoint closed universe component was constructed. The corresponding boundary state is a low temperature partially entangled thermal state (PETS) below the Hawking-Page threshold. This provides a concrete realization of a closed universe encoded in an ordinary holographic setup.\footnote{Other bulk realizations of closed universes are proposed, for example in \cite{Maldacena:2004rf,Freivogel:2005qh,Balasubramanian:2023xyd,Mirbabayi:2020grb}. } The very existence of this geometry raises an interesting puzzle, first pointed out in \cite{Antonini:2024mci}. Since this bulk spacetime is dual to a PETS \cite{Goel:2018ubv} constructed by applying light defect operators to a thermofield double state below the Hawking-Page threshold, it seems that there should also be another bulk description: two disjoint AdS spacetimes with an entangled state of bulk quantum fields, but without any closed universe. Several ideas for its resolution have been proposed since then \cite{Engelhardt:2025vsp,Engelhardt:2025azi,Gesteau:2025obm}. Also PETSs in SYK  and their dual two dimensional baby universes are studied recently \cite{Sasieta:2025vck,Sontag:2026iiu}.

In \cite{Liu:2025cml} and \cite{Kudler-Flam:2025cki}, it has been argued that some semiclassical closed universes in bulk geometry admit a holographic CFT description in terms of states that do not have a well defined large $N$ limit in the usual sense (i.e., states that do not allow norm convergence in the limit). For this result, \cite{Liu:2025cml} and \cite{Kudler-Flam:2025cki} pointed out that although the states themselves do not converge, some of their expectation values as well as correlation functions do converge. These convergent quantities see a mixed state in the large $N$ limit, signaling the emergence of a closed universe in the bulk geometry.

Clearly, a concrete construction of such a class of states on the CFT side is highly desirable. This is challenging because one must start with a finite $N$ theory, select a sequence of states, and then carefully track how these states change as $N$ is increased. A two dimensional symmetric orbifold theory is an exception, providing a tractable example that allows a detailed study of the nature of the large $N$ limit.

In this paper, we study such a class of states in symmetric orbifold CFTs. We focus on a subspace of the theory obtained by taking $N$ copies of a low energy subspace of the seed theory, and then imposing the orbifold gauge constraint associated with the permutation group $S_N$.

We will show that, before imposing the $S_N$ gauge constraint, the large $N$ Hilbert space decomposes into superselection sectors labeled by occupation number distributions. Among these sectors, the largest sector naturally exhibits several features expected of a macroscopic closed universe sector. One reason is that the maximally mixed state is accommodated in this sector, and by the correlation functions of simple operators each pure state in this sector is indistinguishable from this maximally mixed state. As argued in \cite{Liu:2025cml,Kudler-Flam:2025cki}, 
this indistinguishability between pure states and mixed states as probed by large-$N$ simple correlators, is precisely the condition that allows us to interpret this sector as being dual to a closed universe in the bulk.

Furthermore, we will show that this sector admits a natural action of a type II$_1$ von Neumann algebra. Indeed, in the strict large $N$ limit,  the maximally entropic sector corresponds to the infinite tensor product of finite dimensional Hilbert spaces equipped with the maximally mixed state. This is precisely the standard construction of the hyperfinite type II$_1$ factor. (See e.g., \cite{Witten:2018zxz,Liu:2025krl} for a review on the construction.) All of these results appear to imply that the maximally entropic sector in the ungauged Hilbert space is dual to a de Sitter like closed universe in the bulk.

We then impose the orbifold gauge constraint and show that the large permutation symmetry drastically reduces the number of independent states. The basic reason is simple: the dimension of the original ungauged subspace is $k^N$, where $k$ is the number of low energy seed states we keep, whereas the order of the gauge group is $|S_N|=N!\sim N^N$. Thus, for fixed $k=O(1)$, almost all states in the ungauged Hilbert space are in the same orbit of the group action. This provides a CFT realization of the mechanism by which a seemingly large closed universe Hilbert space can become  one dimensional at leading semiclassical order. The reduction in the number of physical states in the presence of permutation symmetry has been studied previously in the context of the black hole interior, for example in \cite{Maxfield:2023mdj}

However, not all states lie in the same orbit. Two ungauged states with different occupation numbers cannot be related by the action of $S_N$. Therefore, after gauging, one independent state remains for each occupation number sector. Since the number of such sectors grows as
\be
\binom{N+k-1}{k-1}\sim N^{k-1},
\ee
the gauged Hilbert space still contains polynomially many degrees of freedom.

Furthermore, each state in the gauged Hilbert space belongs to a distinct superselection sector labeled by its occupation number distribution. This leads to the following picture: The original ungauged Hilbert space decomposes into many superselection sectors corresponding to different baby universes in the bulk, and gauging by $S_N$ collapses each sector to a single physical state in  leading order in the large $N$ expansion. The largest de Sitter like universe is no exception.
This reproduces the conclusion obtained from the bulk gravitational path integral with the full inclusion of wormhole contributions \cite{Harlow:2025pvj}.
For further studies of the relation between wormholes and superselection sectors, we refer to \cite{Marolf:2020xie, Iizuka:2021tut,Coleman:1988cy,Giddings:1988cx,Giddings:1988wv}.

Our CFT analysis suggests that using the dominant saddle approximation for a gravitational path integral $Z_{\rm gravity}$ involving closed universes in the bulk may sometimes be in tension with the bulk to boundary dictionary. This is because, on the gravity side, the saddle point approximation only keeps the contribution of the largest closed universe sector. On the CFT side, however, all superselection sectors contribute to the corresponding quantity at the same magnitude.

This paper is organized as follows. In section \ref{sec:orbifold}, we review symmetric orbifold CFTs and their holographic interpretation. In section \ref{sec:state}, we construct non convergent large $N$ sequences of states from a low energy subspace of the seed theory, classify them by occupation number distributions, and show that the resulting sectors behave as superselection sectors for simple operators. In section \ref{sec:hol-inter}, we explain how these sectors suggest a closed universe interpretation in the bulk. In section \ref{sec:vNII1}, we show that the maximally entropic sector naturally gives rise to the hyperfinite type II$_1$ factor. In section \ref{sec:one-dim}, we impose the $S_N$ gauge constraint and show that it reduces each occupation sector to a single gauge invariant state, making the constrained Hilbert space effectively one dimensional at leading exponential order while leaving polynomially many sectors. 
In section \ref{subsec:External}, we introduce observer degrees of freedom and couple them with non gauge invariant states to make the total state gauge invariant. Subsequently, we construct a density matrix for the observer system and evaluate its entanglement entropy.
In section \ref{sec:bulk}, we discuss the bulk interpretation of the surviving polynomial degrees of freedom after imposing the gauge constraint.
In section \ref{sec:conclusion}, we summarize our results and discuss some future directions.
Appendix \ref{app:permutation-averages} contains the combinatorial and Gaussian derivations of the permutation averages used in the main text.
Appendix \ref{app:variance} presents the evaluations of the variances of the norm and overlap under the averaging.

\section{Orbifold theory} 
\label{sec:orbifold}

We first introduce a family of conformal field theories known as symmetric orbifolds, labeled by a positive integer $N$, together with the Hilbert space $\hm_{N}$ for each fixed $N$. We then construct a sequence of states $\{ |\Psi_N\rangle\}_{N},\;|\Psi_N\rangle \in \mathcal{H}_N $, which in general do not converge in the large $N$ limit.

To begin with, we consider a conformal field theory (CFT), referred to as the seed theory $\mathcal{C}_{\rm S}$, with Hilbert space $\mathcal{H}_{\rm S}$. 
The orbifold theory is obtained by taking $N$ copies of the seed theory and quotienting by a finite group $G$ acting on the copies. 
While various choices of $G$ are possible, in this work we focus primarily on the permutation group $S_{N}$. Hereafter we denote this theory $T_{N} =(\mathcal{C}_{\rm S})^{N}/S_{N}$. The central charge of this theory is $Nc_{{\rm seed}}$, where $c_{{\rm seed}}$ is that of the seed theory.

In this paper, we mainly  focus on the untwist sector of orbifold theory. The Hilbert space of this  sector is  given by 
\be
\mathcal{H}^{(N)}_{\rm orb} = (\mathcal{H}_{\rm S})^{N}/S_{N}.
\ee

Unnormalized states in $\mathcal{H}^{(N)}_{\rm orb}$ take the following form:
\be
\ket{\Psi}_{\rm orb}
= \mathrm{Sym}\!\left[ \ket{\phi_1} \otimes \cdots \otimes \ket{\phi_N} \right],
\ee
where $\mathrm{Sym}$ denotes symmetrization with respect to the permutation group $S_N$. 

We can also write the unnormalized state as 
\be
\ket{\Psi}_{\rm orb} = P\ket{\phi_1} \otimes \cdots \otimes \ket{\phi_N},\quad P= \f{1}{|S_{N}|} \sum_{\pi \in S_{N}} U(\pi),
\ee 
where $U(\pi)$ acts on the states as 
\be
U(\pi)\ket{\phi_1} \otimes \cdots \otimes \ket{\phi_N} =\ket{\phi_{\pi(1)}} \otimes \cdots \otimes \ket{\phi_{\pi(N)}}.  \label{eq;action}
\ee

Similarly, operators on $\mathcal{H}_{\rm orb}$ can be constructed from operators in the seed theory,
\be
\mathbf{O}= \mathrm{Sym}\!\left[ O_{1} \otimes O_{2} \otimes \cdots O_{N} \right] =\sum_{\pi \in S_{N}} U(\pi)\left[ O_{1} \otimes O_{2} \otimes \cdots O_{N} \right] U^{\dagger} (\pi).
\ee \label{eq;defop}
\subsection{Relation to holography}

Symmetric orbifold theories have been studied extensively in the context of holography. For example, the D1-D5 CFT at a particular point in its moduli space,  is described by this class of CFTs \cite{Strominger:1996sh}. It has been shown that its holographic dual is not well described by ordinary Einstein gravity, but instead involves a tensionless string whose string length is comparable to the AdS radius. For instance, in the large $N$ limit, the density of states exhibits Hagedorn growth \cite{Belin:2014fna} below the black hole threshold. This result is universal in the sense that it holds for any choice of the seed theory. See \cite{Belin:2015hwa,Haehl:2014yla,Keller:2017rtk,Belin:2025nqd} for recent related studies. In some particular cases, the exact holographic duality has been derived \cite{Eberhardt:2018ouy,Eberhardt:2019ywk}.

In this theory  the ground state is given by the product of the seed theory vacuum $|0\ra$,  i.e.,$\;|0 \ra_{\rm orb} =|0\ra^{\otimes N}$  and this is dual to the global AdS$_{3}$. Also the operators  \eqref{eq;defop} with a few non identity operators are dual to bulk QFT operators on the fixed AdS$_{3}$ background.

\section{Non convergent sequences of states in orbifold theories }\label{sec:state}

Let $\mathcal{L}$ be the set  consists of first $k$ states $\{\ket{x_{1}},\;\ket{x_{2}},\cdots \ket{x_{k}}\} $
in the spectrum of the seed theory from the lowest energy. We choose $k$ to be $O(1)$ when  taking the large $N$ limit. We then construct a sequence of states with length $N$ $\;\big(\ket{x_{i_{1}}},\;\ket{x_{i_{2}}},\cdots \ket{x_{i_{N}}} \big)$ drawn from $\ml$, and form a state $\ket{\Psi_{N}}_{{\rm orb}} $ in the rank $N$ orbifold theory $T_{N}$, 

\be
\big(\ket{x_{i_{1}}},\;\ket{x_{i_{2}}},\cdots \ket{x_{i_{N}}} \big)\rightarrow   
\ket{\Psi_{N}}_{{\rm orb}} = \mathrm{Sym}\!\left[ \ket{x_{i_{1}}} \otimes\ket{x_{i_{2}}} \cdots \otimes \ket{x_{i_{N}}} \right]. \label{eq;orbstate}
\ee

We can  construct a state $\ket{\Psi_{N+1}}_{{\rm orb}}$ in the rank $N+1$ theory $T_{N+1}$,  from $\ket{\Psi_{N}}_{{\rm orb}}$ in $T_{N}$ by adding one state in $\ml$. Since there are $k$ states in $\ml$, there are $k$ possible choices for $\ket{\Psi_{N}}_{{\rm orb}}$. As we increase the rank $N$, say $N+N_{1}$ there are more and more possible states $\ket{\Psi_{N+N_{1}}}_{{\rm orb}}$
obtained by starting from $\ket{\Psi_{N}}_{{\rm orb}}$.  This means that, if we consider the sequence of the states $\{|\Psi_{N} \ra_{{\rm orb}}\}$,  it does not converge in the large $N$ limit.

The nature of these states is different from that of black hole microstates in the orbifold theory. Calculations of two point functions \cite{Balasubramanian:2005qu,Belin:2025nqd} for example
show that typical black hole microstates in this theory belong to the twisted sector. 
By contrast, the states discussed in this work lie in the untwisted sector, so that they remain below the black hole threshold.

\subsection{Defining a probability distributions from non convergent states.}

 Although the sequence we constructed does not have a well defined large $N$ limit, these states can still be classified in the large $N$ limit by assigning a state dependent probability distribution. For fixed $N$ and a state $\ket{\Psi_N}_{\rm orb} $ of the form \eqref{eq;orbstate}, let $N_{\Psi}(x_i)$ denote the number of times that the state $\ket{x_i}\in \ml$  appears in $\ket{\Psi_N}_{\rm orb}$. The probability $p^{N}_{\Psi} (x_{i})$ of finding $\ket{x_i} \in \ml$ in $\ket{\Psi_N}_{\rm orb}$ is 
\be
p^{N}_{\Psi} (x_{i})= \f{N_{\Psi}(x_i)}{N}, \quad i=1,\cdots k \label{eq;defprob}.
\ee

For each {\it fixed} sequence, this probability distribution has a natural large $N$ limit,
\be
p(x_{i}) =\lim_{N \rightarrow \infty} p^{N}_{\Psi}(x_{i} ). \label{eq:largeNprob}
\ee
Hereafter we denote $\vec{p}=\big ( p(x_{1}), \cdots p(x_{k}) \big) $.

\subsection{The Hilbert space of non gauge invariant states and observers} \label{sec;prob}

We then define a sub Hilbert space $\hm[\vec{p}]$  in the large $N$ limit of $\mathcal{H}^{(N)}_{\rm orb}$ from the probability distribution $\vec{p}$,  spanned by the states that generate $\vec{p}$ through \eqref{eq;defprob}.

Furthermore, we have to imposed the gauge constraint. We will do so in the sections later, and let us first discuss the structure of the Hilbert space {\it without} the gauge constraint in this section.

One might wonder why we discuss non gauge invariant sectors at all. Since gauge degrees of freedom are redundancies of the description, they are not expected to appear as independent degrees of freedom in the physical Hilbert space, and hence do not by themselves have a direct gravitational dual.

There is, however, a useful sense in which such non singlet sectors can become relevant once the system is coupled to an external source. As discussed in \cite{Chandrasekaran:2022cip}, a closed universe with a nontrivial Hilbert space should be understood not as an isolated gauge invariant subsystem, but as a sector that becomes gauge invariant only after being combined with an external system. In other words, the closed universe sector may carry nontrivial gauge charge by itself, while the total system, consisting of the closed universe together with the external source, is gauge invariant.
We mention the  construction of this bipartite system in detail in section \ref{subsec:External}.

In the gravitational interpretation, this external source is naturally identified with the observer degrees of freedom. Thus the non gauge invariant sectors considered above should not be regarded as physical states of the original closed system by themselves. Rather, they provide an auxiliary description of the closed universe Hilbert space that becomes physical once the appropriate observer degrees of freedom are included.

Let $\hm_{\ml}^{\otimes N}$ be the large $N$ Hilbert space obtained by tensoring states chosen from the set $\ml$ of the low energy states in the seed theory, i.e., ${\rm dim }\;\hm_{\ml}=k$, then it is  naturally regarded as the direct sum of $\hm[\vec{p}]$,
\be
\hm_{\ml}^{\otimes N} = \bigoplus_{\vec{p}}\; \hm[\vec{p}].
\ee

Each of these vector spaces constructed in this way form a superselection sector with respect to low energy operators with $O(1)$ conformal dimension. For instance, let us consider a state $|\Psi_{\vec{p}} \ra= \ket{x_{i_{1}}} \otimes\ket{x_{i_{2}}} \cdots \otimes \ket{x_{i_{N}}} $ in $\hm[\vec{p}]$, and excitations on top of it. Such an excitation is obtained by acting an operator of the form 
\be
\mathbf{O}= {\rm Sym} \big[1 \otimes  \cdots O_{i_{1}}\otimes 1  \otimes \cdots O_{i_{2}} \cdots \otimes \cdots  \big],\label{eq:orbifoldop}
\ee
where $O_{i_{k}}$ denotes a non identity operator in the seed theory. The number of non trivial operators in the right hand side of \eqref{eq:orbifoldop} does not scale with $N$ in order for $\mathbf{O}$ to be a low enegy operator. Holographically, this condition implies that $\mathbf{O}$ is dual to an operator in the bulk effective quantum field theory on a fixed background.  In what follows, we call the class of operators simple operators. 

We act this operator $\mathbf{O}$ to a state $|\Psi_{\vec{p}} \ra$ in the Hilbert space $\hm[\vec{p}]$
of our interest, then again compute the probability of finding $|x_{i}\ra$ in $\mathbf{O}|\Psi_{\vec{p}} \ra \,$, namely, $\;p_{\mathbf{O}}(x_{i}) \equiv \lim_{N \rightarrow \infty} \;p^{N}_{\mathbf{O}\Psi}(x_{i} )$  using the formula \eqref{eq;defprob}. Since the number of the nontrivial operators in $\mathbf{O}$ is finite, the probability distribution is unchanged, i.e., $p_{\mathbf{O}}(x_{i}) =p(x_{i})$. This means that $\mathbf{O}|\Psi_{\vec{p}} \ra$ is still  in $\hm[\vec{p}]$. This in particular means that  a state in 
$\hm[\vec{p}]$ can not be mapped to the state with other probability distribution by a low energy operator, therefore  $\hm[\vec{p}]$ forms a superselection sector.

We can determine the dimension of such a Hilbert space. Since the number of possible tensor product states yielding $\vec{p}$ is, in the large $N$ limit, characterized by the von Neumann entropy $S(\vec{p})$ of the probability distribution, we have
\be
{\rm dim} \hm[\vec{p}] =\exp\left[- N \sum_{i} p_{i} \log p_{i} \right] =e^{N S(\vec{p} )}. \label{eq;dim}
\ee

This implies that among all such vector spaces determined by a probability distribution, the largest one is given by $\vec{p} =\vec{p}_{\rm max}$, where 
\be
{p}_{\rm max} (x_{i}) = \f{1}{k} \quad i=1, \cdots k.
\ee

\subsection{A pure state in$\;\hm[\vec{p}]\;$ is indistinguishable from a mixed state }
Now we argue that at the level of the correlation functions of simple operators \eqref{eq:orbifoldop} having a well defined large $N$ limit,  a pure state $|\Psi_{N} \ra$ in the large $N$ Hilbert space $\hm[\vec{p}]$ is indistinguishable from a mixed state $\rho_{\infty}$ defined by the probability distribution $\vec{p}$,
\be 
 \lim_{N \rightarrow \infty}\la \Psi_{N}  | \mathbf{O}_{1} \cdots \mathbf{O}_{m}|\Psi_{N} \ra = {\rm tr}[\rho_{\infty}\mathbf{O}_{1} \cdots \mathbf{O}_{m}  ], \label{eq;indist}
\ee
where the number of operator insertions $m$ is finite and does not scale with $N$\footnote{See, e.g., \cite{Balasubramanian:2005mg} for a similar discussion in black hole contexts.}.  
This mixed state  $\rho_{\infty}$ is obtained as the large $N$ limit of the mixed state  $\rho^{\otimes N} $, where  $\rho$ is the density matrix of the seed theory  on the sub Hilbert space $\hm_{\mathcal{L}}$ spanned by the set of states $\mathcal{L}$,
\be
\rho= \sum_{i=1}^{k} p(x_{i}) |x_{i} \ra \la x_{i}|.
\ee

The state in the left hand side of \eqref{eq;indist} has the form
\be
| \Psi_{N} \ra =|x_{i_{1}} \ra  |x_{i_{2}} \ra  \cdots |x_{i_{N}} \ra , \quad 1 \leq x_{i_{l}} \leq k .\quad 
\label{eq;unconst}
\ee
We stress again that, at this stage, we do not impose the gauge constraint. 
 Let us compute simplest correlation function, namely the one point function of the operator 
\be
\mathbf{O}=  {\rm Sym} \left[ O \otimes 1 \otimes 1 \cdots 1 \ \right] = \f{1}{N}\sum_{m=1}^{N} 1^{\otimes m-1} \otimes O \otimes   1^{\otimes N-m}
\ee
with the seed theory operator $O$,
\be 
\la \Psi_{N} | \mathbf{O} | \Psi_{N} \ra = \f{1}{N}\sum_{l=1}^{N} \la x_{i_{l}} |O |x_{i_{l}} \ra \rightarrow  \sum_{i=1}^{k} p(x_{i}) \la x_{i}| O|  x_{i} \ra = {\rm tr}_{\hm_{\ml}^{\otimes N}} [\rho^{\otimes N}  \mathbf{O} ], \quad N \rightarrow \infty.
\ee
In the second line we used the fact that the probability \eqref{eq;defprob}  approaches \eqref{eq:largeNprob}. This argument is trivially generalized to $m$  point correlations function  as long as $m \ll N $. 

This story is precisely what was argued in  \cite{Liu:2025cml,Kudler-Flam:2025cki}, namely a convergent correlation function of a non convergent state may see a mixed state. Also it is worth pointing out that the way we obtained the maximally mixed state is reminiscent of the quantum de Finetti theorem. For its relations to black hole information loss problem, see \cite{Renner:2021qbe}.

\section{Holographic interpretation of the unconstrained Hilbert space}\label{sec:hol-inter}

We have seen that the ungauged Hilbert space $\hm[\vec{p}]$ contains a maximally entropic state and at the level of correlation functions of simple operators of the form \eqref{eq:orbifoldop},  a pure state  in this Hilbert space is indistinguishable from the mixed state $\rho_{\infty} = \lim_{N \rightarrow \infty} \rho^{\otimes N}$, as in \eqref{eq;indist}. This signals that its dual gravity description involves closed universes in the bulk, as advocated in \cite{Liu:2025cml,Kudler-Flam:2025cki}.

For instance, in the dual gravity description, the simple operators correspond to the operators in the bulk quantum field theory on the region connected to the asymptotic boundary, called causal wedge. $\rho_{\infty}$ is dual to the bulk density matrix $\rho_{{\rm bulk}}$ on the causal wedge, and furthermore, it is a mixed state. Since the original CFT state in $|\Psi_{N} \ra$ is pure, the total bulk state must also be pure, this implies that there must be an additional bulk component in addition to the causal wedge, which purifies the bulk density matrix. This additional bulk component is identified with a closed baby universe. 

We also would like to point out that  the fact that the class of  the Hilbert spaces $\hm[\vec{p}]$ parametrised by a probability distribution form superselection sector which   is also one of the  basic properties of a baby universe.  

Below we strengthen this proposal  by showing the von Neumann algebra acting on the largest Hilbert space $\hm[\vec{p}_{\rm max}]$ is of type II$_{1}$ in section \ref{sec:vNII1}.

\section{Hyperfinite type II$_1$ factor from $\hm[\vec p_{\rm max}]$}
\label{sec:vNII1}

In this section, we demonstrate that the von Neumann algebra associated with the largest sector $\hm[\vec p_{\rm max}]$ is the hyperfinite type II$_1$ factor.
Mathematically, this construction corresponds to an infinite tensor product of finite type I factors (ITPFI) \cite{Araki1968ACO}. We summarize only the essential elements relevant to our physical setup; we refer the reader to \cite{Witten:2018zxz,Sorce:2023fdx,Liu:2025krl} for further details.

While our primary interest lies in the largest sector with $\vec p_{\rm max}$, we begin by formulating the construction for a generic probability distribution. For a fixed rank $N$, the probability distribution introduced in section \ref{sec:state} yields the mixed state
\begin{equation}
\rho_{\vec p_\Psi^N} = \sum_{i=1}^{k} p_\Psi^N(x_i) \ket{x_i} \bra{x_i},
\label{eq:seed-mixed-state-Large_N-limit}
\end{equation}
and the corresponding product state is given by
\begin{equation}
\sigma_N = (\rho_{\vec p_\Psi^N})^{\otimes N}.
\label{eq:finiteN-tensor-mixed}
\end{equation}

To isolate the large $N$ probability distribution from the infinite tensor product structure, it is convenient to introduce
\begin{equation}
\tilde\sigma_{N:M} = (\rho_{\vec p_\Psi^N})^{\otimes M},
\end{equation}
where $N$ and $M$ are treated as independent parameters. We first take the limit $N\to\infty$, which yields
\begin{equation}
\tilde\sigma_{\infty:M} = (\rho_{\vec p})^{\otimes M},
\qquad
\rho_{\vec p} = \sum_{i=1}^{k} p(x_i) \ket{x_i} \bra{x_i},
\end{equation}
and subsequently construct the infinite tensor product algebra by taking $M\to\infty$.

For each copy $\nu$, let $\mathcal H_\nu$ denote the $k$ dimensional Hilbert space spanned by the low energy states in $\mathcal L$, and let
\begin{equation}
\mathcal M_\nu = \mathcal B(\mathcal H_\nu)
\end{equation}
be the local type I$_k$ factor. The corresponding state functional is defined as
\begin{equation}
\omega_\nu(a_\nu) = \tr(\rho_{\vec p}\,a_\nu).
\end{equation}
For a finite $M$, we have
\begin{equation}
\mathcal M^{(M)} = \bigotimes_{\nu=1}^{M}\mathcal M_\nu,
\qquad
\omega^{(M)}(a_1\otimes\cdots\otimes a_M) = \prod_{\nu=1}^{M}\omega_\nu(a_\nu).
\end{equation}
As long as $M$ remains finite, this simply describes a standard type I algebra. The non trivial structure emerges only after taking the weak closure in the infinite tensor product limit.

A crucial point is that the probability distribution $\vec p$ determines which infinite sequences of local operators possess well defined expectation values in the limit state.
Equivalently, $\vec p$ specifies the notion of convergence in the large $M$ limit. Consequently, different probability distributions yield distinct classes of allowed limit operators, thereby leading to different von Neumann algebras.

Let us now focus on the largest sector\footnote{One can also consider non-largest sectors. For such cases, one would find that resulting algebras become type I or III factors, by following, e.g., Lemma 2.14 of \cite{Araki1968ACO}. However, since they are sub dominant sectors, we do not discuss them in this paper. It would be interesting to consider corresponding bulk structures and investigate its relation to their algebraic structures.
}, given by
\begin{equation}
\vec p_{\rm max} = \left(\frac1k,\cdots,\frac1k\right),
\qquad
\rho_{\vec p_{\rm max}} = \frac1k I_k .
\end{equation}

The physically relevant observables are the finitely supported operators discussed in section \ref{sec:state},
\begin{equation}
a = a^{(M_0)}\otimes I\otimes I\otimes\cdots = a_1\otimes\cdots\otimes a_{M_0}\otimes I\otimes I\otimes\cdots,
\label{eq:finitely-supported-op}
\end{equation}
which act non trivially only on a finite number of copies. These represent simple low energy operators dual to bulk effective QFT operators on a fixed background.
Allowing infinitely supported operators would generically alter the macroscopic probability distribution and therefore fall outside the low energy sector described by effective bulk QFT.

For such finitely supported operators, the limit state evaluates to
\begin{equation}
\omega_{\rm max}(a) = \tr\left[ \left(\frac1k I_k\right)^{\otimes M_0} a^{(M_0)} \right] = \prod_{\nu=1}^{M_0}\frac1k\tr(a_\nu).
\end{equation}
Thus, the expectation value reduces to the normalized matrix trace over the finite support. In particular, we obtain
\begin{equation}
\omega_{\rm max}(ab) = \omega_{\rm max}(ba)
\end{equation}
for any finitely supported operators $a$ and $b$. This tracial property is preserved upon taking the weak closure. 
Following the standard operator algebraic construction based on finitely supported operators (see, e.g., \cite{Kang:2019dfi}), the weak closure of this algebra defines an infinite dimensional von Neumann factor equipped with a faithful normalized trace, namely a type II$_1$ factor. Furthermore, since it is constructed as the weak closure of an increasing sequence of finite dimensional matrix algebras, it is hyperfinite. See, e.g., \cite{Witten:2018zxz} for an analogous algebraic construction.

Intuitively, the hyperfinite type II$_1$ factor emerges because the infinite tensor product limit preserves the local matrix algebra structure of finitely supported operators, while simultaneously enlarging the algebra through weak limits of operator sequences.

We next clarify the Hilbert space on which this algebra acts, as well as the physical origin of its commutant. The relevant Hilbert space is not the naive infinite tensor product $\bigotimes_\nu\mathcal H_\nu$, but rather the GNS Hilbert space associated with the tracial state $\omega_{\rm max}$. Equivalently, one can purify the maximally mixed single copy state by introducing an auxiliary copy $A$,
\begin{equation}
|\Omega_1\rangle = \frac1{\sqrt k}\sum_{j=1}^{k} |j\rangle_S\otimes|j\rangle_A ,
\label{eq:single-purification}
\end{equation}
where $|j\rangle_S$ denotes a low-energy state of the original system.
The reference state for the infinite system then becomes
\begin{equation}
|\Omega_{\rm max}\rangle = \bigotimes_{\nu=1}^{\infty}|\Omega_\nu\rangle .
\end{equation}
The GNS Hilbert space is constructed by acting with finitely supported operators on this reference state and taking the completion.

In this representation, the observable algebra $\mathcal M$ acts on the original system $S$, while the commutant $\mathcal M'$ acts on the auxiliary system $A$. For finitely supported operators, this implies the simple commutation relation
\begin{equation}
[a_S\otimes I_A,\, I_S\otimes b_A] = 0 .
\end{equation}
In the GNS representation associated with $\omega_{\rm max}$, the commutant algebra $\mathcal M'$ is again a hyperfinite type II$_1$ factor.

We emphasize that these auxiliary degrees of freedom are not introduced artificially by hand. Once the large $N$ limit necessitates that simple observables probe the maximally mixed state $\rho_{\vec p_{\rm max}}$, its canonical purification and equivalently, the commutant algebra automatically emerges. Within our framework, we naturally identify this commutant algebra with the closed universe degrees of freedom.

This doubled structure can also be understood from an algebraic perspective. By regarding the space of operators as a Hilbert space equipped with the Hilbert Schmidt inner product, the algebra acts via left multiplication,
\begin{equation}
\pi_L(a)x = ax,
\end{equation}
while its commutant acts via right multiplication,
\begin{equation}
\pi_R(b)x = xb^T .
\end{equation}
The associativity of matrix multiplication guarantees that these two actions commute. This algebraic viewpoint makes it explicit that the closed universe algebra is intrinsically encoded within the operator algebraic structure as soon as the large $N$ probability distribution is specified.

\section{The constrained Hilbert space in the large $N$ limit}\label{sec:one-dim}

So far we have not imposed the gauge constraint on the Hilbert space. Now we would like to ask what happens once we impose the $S_N$ gauge constraint. In this section, we argue that the gauge constraint drastically reduces the dimension of the Hilbert space, from $e^{N S[\vec{p}]}$ to a quantity that grows only polynomially with $N$. In particular, this means that if we evaluate the dimension in the large $N$ limit by keeping only the leading term in the large $N$ expansion, the Hilbert space of each superselection sector is one dimensional, including the largest one $\hm[\vec{p}_{\rm max}]$.  We argue this is precisely the CFT dual of the fact that the Hilbert space of closed universe is one dimensional.

A basis state on the unconstrained space takes the form \eqref{eq;unconst}.  These states are projected to the  state
\be
|\Psi_{(x_{1}, \cdots ,x_{N})} \ra =\frac{1}{\sqrt{\mathcal{M}}} P|x_{1}, \cdots ,x_{N}\ra =\frac{1}{\sqrt{\mathcal{M}}}\cdot \frac{1}{|S_{N}|}  \sum_{\pi \in S_{N}} |x_{\pi(1)}, \cdots ,x_{\pi(N)}\ra, 
\ee
where $|S_N|$ is the number of elements of the group $S_N$, and $\mathcal{M}$ is the normalization factor.

Notice that even if  we take  two orthogonal  states $|x_{1}, \cdots ,x_{N}\ra, $ $|y_{1}, \cdots ,y_{N}\ra $ in the unconstrained space, the projected states $ |\Psi_{(x_{1}, \cdots ,x_{N})} \ra,|\Psi_{(y_{1}, \cdots ,y_{N})} \ra$ are in general not orthogonal.
Thus, we would like to compute the overlap between two such gauge invariant states.

For convenience, we introduce the vector notation $\vec{x}=(x_1,\ldots,x_N)$ for  each sequence and write the projected state as $\ket{\Psi_{\vec{x}}}\coloneqq \ket{\Psi_{(x_{1}, \cdots ,x_{N})} } $.
Let us first determine the normalization of the state,
\be
\la \Psi_{\vec{x}} 
|\Psi_{\vec{x}} \ra = \frac{1}{\mathcal{M}}\cdot \f{1}{|S_{N}|} \sum_{\pi \in S_{N}} \la x_{1}\cdots x_{N} |x_{\pi(1)}, \cdots ,x_{\pi(N)}\ra. 
\ee
It is hard to evaluate it for a {\it specific} state. Therefore, we instead compute 
its average\footnote{One can also consider normalization factors for specific states. However, to simplify following discussions, we consider the normalization under the average.}, 
\begin{align}
\overline{\;\langle \Psi_{\vec{x}} 
| \Psi_{\vec{x}} \rangle}&=\sum_{\vec{x}=1}^{k} p(\vec{x})\la \Psi_{\vec{x}} 
|\Psi_{\vec{x}} \ra \\
&=\frac{1}{\mathcal{M}}\cdot\f{1}{|S_{N}|k^{N}} \sum_{x_{1}, \cdots ,x_{N}=1}^{k} \sum_{\pi \in S_{N}}\; \prod_{a=1}^{k} \delta_{x_{a},x_{\pi(a)}}. \label{eq;normave}
\end{align}
Notice that each state in the sequence is independently generated from the probability distribution $\vec{p}_{{\rm max}}$, so $p(\vec{x})=p(x_{1}, \cdots x_{N})=1/k^{N}$.

To discuss the specific state in the gauged Hilbert space, we need to calculate how many ungauged states can be projected to the gauged state. In the gauged Hilbert space, states are distinguished by the number of each element in the sequence $\vec{x}=(x_1,\cdots, x_N)$. Let $N_{m}$ be the number of the elements $x_{l}, 1\leq l\leq N\;$in the sequence $\vec{x}$ satisfying $x_{l}=m\; (1\leq m \leq k)$. 
Then sum over sequence is equivalent to sum over these numbers $(N_{1}, \cdots N_{k})$ with $N_{1}+ \cdots N_{k}=N$. The number of sequences with this occupation number is 
\be
\f{N!}{
\prod_{b=1}^{k} N_{b}!  }.\label{eq:dim-occu}
\ee
Therefore, for any function of the sequence $I(x_{1}, \cdots ,x_{N})$, we have 
\be
\sum_{x_{1}, \cdots ,x_{N}=1}^{k} I(x_{1}, \cdots ,x_{N}) = \sum_{\substack{N_{1}, \cdots N_{k} \\ N_{1}+\cdots +N_{k}=N} } I(N_{1}, \cdots ,N_{k}) \; \f{N!}{
\prod_{b=1}^{k} N_{b}!  },
\label{eq;occup}
\ee
where the $I(N_1,\cdots,N_k)$ is a function of the occupation numbers.

By writing the summand of \eqref{eq;normave} by occupation numbers, and plugging it to the formula \eqref{eq;occup}, we obtain\footnote{In appendix \ref{app:permutation-averages}, we give another derivation of this result using Gaussian integrals.}
\be
\overline{\;\langle \Psi_{\vec{x}} 
| \Psi_{\vec{x}} \rangle} = \frac{1}{\mathcal{M}}\cdot \f{1}{|S_{N}|k^{N}}
\left( \f{(N+k-1)!}{(k-1)!} \right)=\frac{1}{\mathcal{M}}\cdot\frac{1}{k^{N}} \binom{N+k-1}{N},\label{eq:norm-finite-N}
\ee
where, in the last equality, we used $|S_{N}|=N!$. Thus, the normalization factor $\mathcal{M}$ is given by
\begin{equation}
    \mathcal{M}=\frac{1}{k^{N}} \binom{N+k-1}{N}=\frac{C_{N,k}}{k^{N}}.\label{eq:normalization-factor-M}
\end{equation}
Here, we introduced
\begin{equation}
    C_{N,k}\coloneqq \binom{N+k-1}{N}. \label{eq:bi-nomial}
\end{equation}

Similarly, we can compute the average of overlaps between two states 
$\overline{\;\langle \Psi_{\vec{y}} 
| \Psi_{\vec{x}} \rangle}$\footnote{In appendix \ref{app:permutation-averages}, we give the derivation of this result.},
\be
\begin{aligned}
    \overline{\;\langle \Psi_{\vec{y}} 
| \Psi_{\vec{x}} \rangle}
&=\frac{1}{\mathcal{M}}\cdot \left(\f{1}{|S_{N}|\;k^{2N}} \sum_{\pi \in S_{N}} \sum_{x_{1}, \cdots ,x_{N}=1}^{k} \sum_{y_{1}, \cdots ,y_{N}=1}^{k}  \prod_{a=1}^{N} \delta_{y_{a} x_{\pi(a)}} \right)\\
&=\frac{1}{\mathcal{M}}\cdot \left( \f{1}{k^{N}}\right)\\
&=\frac{1}{C_{N,k}},\label{eq:finite-N-overlap}
\end{aligned}
\ee
where we used \eqref{eq:normalization-factor-M}.
We also computed the variances of the norm $\langle \Psi_{\vec{x}} 
| \Psi_{\vec{x}}\rangle$ and the overlap $\langle \Psi_{\vec{x}} 
| \Psi_{\vec{y}}\rangle$ under the average in appendix \ref{app:variance}.

\subsection{Deriving  one dimensional Hilbert space in a particular large $N$ limit}

Here we would like to employ  the particular large $N$ approximation scheme, namely writing quantities of interest in the exponential form $e^{f(N)}$ and picking up only the large $N$ leading term in $f(N)$. This approximation scheme is motivated by the fact that  when we compute a quantity by a semiclassical gravitational path integral in the saddle point approximation, we only keep the leading term in the small $G_{N}$ expansion in the exponent, namely 
\be 
\la X \ra =\int \mathcal{D} g_{\mu \nu} \;  X\;e^{-S_{{\rm gravity}}} \sim e^{-\f{a_{0}}{G_{N}}  + a_{1}\log G_{N}+ \sum_{m} a_{m} (G_{N})^{m} } \sim e^{-\f{a_{0}}{G_{N}} }, \quad G_{N} \rightarrow 0.
\ee

We would like to perform the similar large $N$ approximation in the CFT side and examine its consequence.
In particular, by computing the average of the  overlap $\overline{\;\langle \Psi_{\vec{y}} 
| \Psi_{\vec{x}} \rangle}$ in the scheme,
the  Hilbert space appears to be one dimensional.

To begin with, let us first consider the normalization of the state in this large $N$ approximation scheme.
This leads to the normalization constant $\mathcal{M}_{\infty}$ different from the naive $N \rightarrow \infty$ limit of  \eqref{eq:normalization-factor-M}.
To this end, we consider the following large $N$ behavior 
\be
\overline{\;\langle \Psi_{\vec{x}} 
| \Psi_{\vec{x}} \rangle}  \rightarrow \frac{1}{\mathcal{M}_{\infty}}\cdot \f{1}{k^{N}} \left( \f{N^{k-1}}{(k-1)!} \right), \quad  N \rightarrow \infty.  \label{eq;expvn}
\ee 

In \eqref{eq;expvn}, if we write it in the exponential form $e^{-N\log k +(k-1)\log N\cdots}$ and only keep the term leading in the large $N$ in the exponent, polynomial part of the right hand side of \eqref{eq;expvn} is replaced by $1$, and as a result   we obtain 
\be
\overline{\;\langle \Psi_{\vec{x}} 
| \Psi_{\vec{x}} \rangle} \approx \frac{1}{\mathcal{M}_{\infty}}\cdot \f{1}{k^{N}}. \label{eq:Leading-N-norm}
\ee
Thus, by picking up a leading contribution in the large $N$ limit, the normalization constant $\mathcal{M}_{\infty}$ is given by
\begin{equation}
    \mathcal{M}_{\infty}=\frac{1}{k^{N}}\label{eq:largeN-normal}.
\end{equation}

Then, we consider the large $N$ leading behavior of the overlap. The factor inside the bracket of the first line in \eqref{eq:finite-N-overlap} is exactly $1/k^{N}$ and does not have sub leading contributions in the exponential form. Thus, it is enough to replace $\mathcal{M}$ with $\mathcal{M}_\infty$, and we obtain
\begin{equation}
    \overline{\;\langle \Psi_{\vec{y}} 
| \Psi_{\vec{x}} \rangle} \approx \frac{1}{\mathcal{M}_{\infty}}\cdot\f{1}{k^{N}}=1,\label{eq:degenerate-Large-N-overlap}
\end{equation}
where, in the last equality, we used \eqref{eq:largeN-normal}.
Therefore, this means that the Hilbert space with the gauge constraint is  one dimensional in this particular large $N$ approximation scheme.

\subsection{An intuitive argument}

Let us explain why we get the one dimensional Hilbert space.  To this end,  we pick up a typical sequence $\vec{x}=(x_{1}, \cdots, x_{N})$ and argue that the gauge orbit of $|x_{1}, \cdots ,x_{N}\ra$ sweeps  all  basis states of the unconstrained space $\hm_{\ml}^{\otimes N}$ with ${\rm dim }\hm_{\ml}^{\otimes N}=k^{N}$ .

Let $G(x_{1}, \cdots x_{N})$ be the subgroup of $S_{N}$  which keep the sequence invariant. Again  by denoting $(N_{1}, \cdots N_{k})$  be the occupation numbers of the sequence $\vec{x}$,  then $G(x_{1}, \cdots x_{N})$  is generated by the smaller permutation groups $S_{N_{1}}, \cdots ,S_{N_{k}}$. Then the coset $S_{N}/G(x_{1}, \cdots x_{N})$ nontrivially acts on the state  $|x_{1}, \cdots ,x_{N}\ra$. When the sequence is typical, $N_{1}= \cdots N_{k} =N/k$, the order of the coset  is 
\be
|S_{N}/G(x_{1}, \cdots x_{N})|= \f{N!}{N_{1}! \cdots N_{k}!} \sim  k^{N}, \quad N \rightarrow \infty.
\ee
This means that gauge orbit of  $|x_{1}, \cdots ,x_{N}\ra$ can generate all basis states. Thus the gauged Hilbert space is effectively one dimensional in the large $N$ limit. 

\subsection{Gram Matrices}

In the previous discussion, we found that the overlap between two projected states becomes of order one if we keep only the leading exponential  contribution in the large $N$ expansion. In this sense, the gauged Hilbert space appears effectively one dimensional at leading semiclassical order. However, this leading approximation discards the polynomially many subleading contributions in $N$.

In this section, we refine this statement by computing the averaged Gram matrix of the projected states. We will show that, the exact gauge invariant Hilbert space contains polynomially many independent directions. More precisely, its dimension grows as $N^{k-1}$. This polynomial degeneracy counts the number of distinct occupation number sectors, which become superselection sectors in the large $N$ limit, therefore we conclude that there is only one state in each superselection sector.  

The reason  why the dimension of the gauged Hilbert space agrees with the number of superselection sectors is simple. Two ungauged states with different occupation numbers cannot be related by the action of the permutation group $S_N$, and hence cannot belong to the same gauge orbit. Then, after imposing the gauge constraint, each occupation number sector contributes one independent gauge invariant state. Since the number of such sectors is $C_{N,k}$, the gauged Hilbert space retains $C_{N,k}$ independent states.

\subsubsection{Quantitative argument using single averaged Gram matrix}

We first rephrase the same argument in the previous subsection from the perspective of the Gram matrix, whose rank quantitatively measures the dimension of the underlying Hilbert space.
To this end, starting from \eqref{eq:norm-finite-N} and \eqref{eq:finite-N-overlap}, we consider the \textit{averaged} Gram matrix given by\footnote{Note that the averaged Gram matrix is slightly different from a non averaged one, which we do not investigate in this paper. }
\begin{equation}
  \overline{G}_{\vec{x}\vec{y}} :=  \overline{\langle \Psi_{\vec{x}}|\Psi_{\vec{y}}\rangle} = \begin{cases} 1, & \vec{x}=\vec{y},\\[2mm]
  \displaystyle \frac{1}{C_{N,k}}, & \vec{x}\neq \vec{y}. \end{cases}
\end{equation}
For notational simplicity, let $J \coloneqq k^{N}$, and let us express it as the matrix form
\begin{equation}
    \overline{G}=\alpha I_J+\beta \mathbf{1}_J,
\end{equation}
where $\alpha\coloneqq 1-1/C_{N,k}$, $\beta\coloneqq 1/C_{N,k}$, $I_J$ denotes the $J\times J$ identity matrix, and $\mathbf{1}_J$ represents the $J\times J$ all ones matrix,
\begin{equation}
    I_J=\mathrm{diag}(1,1,\cdots,1), \qquad  \mathbf{1}_J = \begin{pmatrix}
        1 & 1& \cdots & 1\\
        1 & 1 & \cdots & 1 \\
        \vdots & \vdots & \ddots & \vdots \\
        1 & 1 & \cdots &1 
    \end{pmatrix}.
\end{equation}
Then, the all one vector $\ket{u}=(1,1,\ldots,1)^T$ acts as a  eigenvector satisfying $\overline{G}\ket{u}=(\alpha+\beta J)\ket{u}$, yielding the unique non degenerate eigenvalue
\begin{equation}
  \lambda_1 = 1-\frac{1}{C_{N,k}}+\frac{k^N}{C_{N,k}}.
\end{equation}
Any vector $\ket{v}$ orthogonal to $\ket{u}$ satisfies $\mathbf{1}_J\ket{v}=0$\footnote{Note that $\mathbf{1}_J = \ket{u}\bra{u}$.}, giving the remaining $J-1$ degenerate eigenvalues:
\begin{equation}
  \lambda_2=\cdots=\lambda_J = \alpha = 1-\frac{1}{C_{N,k}}.
\end{equation}
The exact spectral decomposition takes the form $\overline{G} = \lambda_1\ket{e_1}\bra{e_1} + \sum_{a=2}^{J}\lambda_a\ket{e_a}\bra{e_a}$ with $\ket{e_1}\propto \ket{u}$, $\langle e_1|e_b\rangle =0\, (b=2,\cdots,J)$, $\langle e_a|e_b\rangle =\delta_{ab}$. Since all eigenvalues are strictly positive at finite $N$, we obtain the full exponential dimension as our first notion of rank:
\begin{equation}
  \mathrm{rank}\; \overline{G}=J=k^N.
\end{equation}

However, in the large $N$ limit at fixed $k$, where $C_{N,k} \approx N^{k-1}/(k-1)!$, only one eigenvalue diverges:
\begin{equation}
  \begin{gathered}
      \lambda_1 \approx \frac{(k-1)!}{N^{k-1}}k^N, \\
  \lambda_2=\cdots=\lambda_J \approx 1. 
  \end{gathered}
\end{equation}
Keeping only the diverging eigenvalue yields the rank one approximation $\overline{G}_{\mathrm{div}} := \lambda_1\ket{e_1}\bra{e_1}$, giving $\mathrm{rank} \;\overline{G}_{\mathrm{div}}=1$. More precisely, rescaling by the leading eigenvalue gives $\lambda_1^{-1}\overline{G} \to \ket{e_1}\bra{e_1}$ as $N\to\infty$. Every state orthogonal to $\ket{e_1}$ vanishes into the kernel of this limiting inner product, rendering the asymptotic Hilbert space \textit{effectively} one dimensional.

\subsubsection{Refined approach to the rank of averaged Gram matrices}\label{subsec:eff-rank}

The simple averaging performed in the previous subsection would discard terms scale polynomial in $N$. Therefore instead of averaging the Gram matrix directly, we refine the averaging procedure to better probe its rank. Specifically, the  number of basis states can be obtained by evaluating the effective rank from first computing all  $n$-th moments of the Gram matrix, analytically continuing the result in $n$, and then taking the $n\to 0$ limit.  Let $\vec{x}^{(1)}, \dots, \vec{x}^{(n)}$ label $n$ independent copies of the unconstrained configurations under the cyclic identification $\vec{x}^{(n+1)} \equiv \vec{x}^{(1)}$. The average trace of the $n$-th power of the Gram matrix, $\overline{\tr(G^n)} = \sum_{\vec{x}^{(1)}, \dots, \vec{x}^{(n)}} \overline{\langle \Psi_{\vec{x}^{(1)}} | \Psi_{\vec{x}^{(2)}} \rangle \cdots \langle \Psi_{\vec{x}^{(n)}} | \Psi_{\vec{x}^{(1)}} \rangle}$\footnote{We sometimes use the shorthand notation, 
\begin{equation}
    \sum_{\vec{x}}\coloneqq\sum_{x_1,\cdots,x_N=1}^{k}.
\end{equation}
}, maps directly onto the cyclic $n$-copy contraction $\mathsf{F}_{N,k}^{(n)}$ defined in \eqref{eq:Fsf_def}.
This yields an exact identity for any finite $N$ and $k$\footnote{Note that, from the definition of the averaging, we have the following equivalence 
\begin{equation}
    \begin{aligned}
        \overline{\tr(G^n)} &= \sum_{\vec{x}^{(1)}, \dots, \vec{x}^{(n)}} \overline{\langle \Psi_{\vec{x}^{(1)}} | \Psi_{\vec{x}^{(2)}} \rangle \cdots \langle \Psi_{\vec{x}^{(n)}} | \Psi_{\vec{x}^{(1)}} \rangle}\\
        &=\sum_{\vec{x}^{(1)}, \dots, \vec{x}^{(n)}} \langle \Psi_{\vec{x}^{(1)}} | \Psi_{\vec{x}^{(2)}} \rangle \cdots \langle \Psi_{\vec{x}^{(n)}} | \Psi_{\vec{x}^{(1)}} \rangle\\
        &= k^{nN} \cdot \overline{\langle \Psi_{\vec{x}^{(1)}} | \Psi_{\vec{x}^{(2)}} \rangle \cdots \langle \Psi_{\vec{x}^{(n)}} | \Psi_{\vec{x}^{(1)}} \rangle}.
    \end{aligned}
\end{equation}}:
\begin{equation}
  \overline{\tr(G^n)} = \sum_{\vec{x}^{(1)},\dots,\vec{x}^{(n)}} \overline{ \prod_{\alpha=1}^{n} \langle \Psi_{\vec{x}^{(\alpha)}} | \Psi_{\vec{x}^{(\alpha+1)}} \rangle }= k^{nN} \cdot \frac{1}{\mathcal{M}^{n}} \cdot \mathsf{F}_{N,k}^{(n)}= \frac{k^{nN}}{C_{N,k}^{n-1}}, 
\end{equation}
where we used \eqref{eq:Fsf_def}, \eqref{eq:Fsf_res} and \eqref{eq:normalization-factor-M}.

Since $\tr\overline{G}=k^N$, the effective rank can be computed from the normalized $n$-th moment,
\begin{equation}
      \mathrm{rank}_{\mathrm{eff}}^{(n)}(G) \coloneqq \left( \frac{\overline{\tr(G^n)}}{(\tr\overline{G})^n} \right)^{\frac{1}{1-n}} = \left( \frac{k^{nN}/C_{N,k}^{n-1}}{k^{nN}} \right)^{\frac{1}{1-n}} = C_{N,k}\approx \frac{N^{k-1}}{(k-1)!},\label{eq:eff-rank}
\end{equation}
where, in the last equality, we used the large $N$ limit.
Thus, this quantity is completely independent of the index $n$, and this property holds for any finite $N$, indicating that the spectrum of the averaged Gram matrix is perfectly flat across its support.
Then, the $n\to 0$ limit of this quantity is trivial, and the effective rank is given by
\begin{equation}
    \lim_{n\to 0}  \mathrm{rank}_{\mathrm{eff}}^{(n)}(G) =C_{N,k}\approx \frac{N^{k-1}}{(k-1)!},
\end{equation}
where, in the last equality, we again used the large $N$ limit.
This polynomial scale coincides with the number of occupation sectors and therefore captures the number of basis states.

We note that the $n$ independence of the effective rank stems from the relation between the Gram matrix and the projection operator $P$, namely $G_{\vec{x}\vec{y}}=\langle \Psi_{\vec{x}}|\Psi_{\vec{y}}\rangle \propto \langle x_1,\cdots,x_N | P |y_1,\cdots,y_N \rangle$. Therefore, the idempotency of the projection operator ensures the $n$ independence of the effective rank, since $\tr(G^n)\propto \tr(P^{n})=\tr(P)$. From this perspective, we can regard the effective rank as the rank of the projection operator $P$, which determines the dimension of the gauge invariant Hilbert space.

\subsection{ Hilbert spaces for closed universes in CFT}

We have seen that the rank of the averaged Gram matrix of the projected states is proportional to $N^{k-1}$,  which is equal to  the dimension of the Hilbert space with the gauge constraint. To see where these degrees of freedom come from explicitly, it is useful to consider the Gram matrix in the {\it unconstrained} Hilbert space $\hm_{\ml}^{\otimes N}$.  In writing this matrix, it is useful to group the projected states according to their occupation numbers 
\be
\hm_{\ml}^{\otimes N}=\bigoplus_{\alpha}  \hm_{\vN_{\alpha}}, \qquad \vN_{\alpha} \equiv \big(N^{(\alpha)}_{1}, N^{(\alpha)}_{2}, \cdots ,N^{(\alpha)}_{k} \big), \quad N^{(\alpha)}_{1} +\cdots +N^{(\alpha)}_{k}=N.
\label{eq;defsector}
\ee
The dimension of $\hm_{\vN_{\alpha}}$ is $s(\vN_{\alpha}) \equiv N!/N^{^{(\alpha)}}_{1}! \cdots N^{^{(\alpha)}}_{k}!$, which is equal to \eqref{eq:dim-occu}.  One should remark that the large $N$ limit of these Hilbert spaces coincides with the $\hm [\vec{p_{\alpha}}]$ discussed in section \ref{sec;prob} since  $\vec{p_{\alpha}} = \lim_{N \rightarrow \infty}\vec{N_{\alpha}}/N$. Therefore, each $\hm_{\vec{N_{\alpha}}}$ approximately form a superselection sector with respect to simple operators of the form \eqref{eq:orbifoldop}.

Since an overlap between two states with different occupation numbers vanishes, the Gram matrix in the unconstrained Hilbert space take the block diagonal form 

\be
G^{{\rm Ungauged}}= {\rm diag} \Big(G_{\vN_{1}}, G_{\vN_{2}} \cdots \big),
\ee
where $G_{\vN_{\alpha}}$ is the Gram matrix of fixed occupation numbers $\;\vN_{\alpha}$, and it is the $s(\vN_{\alpha})\times s(\vN_{\alpha})$ all ones matrix.
The number of blocks in the Gram matrix is $C_{N,k}$, \eqref{eq:bi-nomial}, therefore $1\leq \alpha \leq C_{N,k}$.

Let us now impose the gauge constraint. After gauging by $S_N$, all states within the same orbit become gauge equivalent.
\begin{equation}
  G^{\mathrm{Ungauged}}
  =
  \begin{pmatrix}
  \begin{tikzpicture}
    \draw node [at={(0.4,-0.4)}]{$G_{\vec{\mathcal{N}_1}}$};
    \draw [dashed] (0.9,0.2)--(0.9,-1.7)--(1.8,-1.7)--(1.8,-0.9)--(-0.3,-0.9);
    \draw node [at={(1.35,-1.35)}]{$G_{\vec{\mathcal{N}_2}}$};
    \draw node [at={(4/2+0.2,-4/2)}]{$\ddots$};
  \end{tikzpicture}
  \end{pmatrix}\;
  \begin{tikzpicture}
    \draw[->] (0,0)--(2,0) node [at={(1,0.5)}]{$S_N$ Gauging};
  \end{tikzpicture}\;
  G^{\mathrm{Gauged}}
  =
  \begin{pmatrix}
  \begin{tikzpicture}
    \draw node [at={(0.1,-0.2)}]{$1$};
    \draw node [at={(0.1+0.45,-0.2-0.4)}]{$1$};
    \draw node [at={(0.1+0.9,-0.2-0.8)}]{$1$};
    \draw node [at={(0.1+1.35,-0.2-1.2)}]{$1$};
    \draw node [at={(0.1+2.0,-0.2-1.6)}]{$\ddots$};
  \end{tikzpicture}
  \end{pmatrix}
  \label{eq;GaugedGram}
\end{equation}
Therefore, each occupation sector contributes only one independent gauge invariant state, independently of the size of the corresponding ungauged block. In other words, gauging replaces each block by a single representative. The number of gauge invariant states is then the number of possible occupation vectors,
\be
\dim \mathcal H_{\rm gauged}
=
C_{N,k}
\sim
\frac{N^{k-1}}{(k-1)!}.
\ee

This is precisely what we have obtained from the calculation of the effective rank of the Gram matrix \eqref{eq:eff-rank}.
In the large $N$ limit with fixed $k$, the largest block in the ungauged Gram matrix corresponds to the sector with approximately uniform occupation numbers,
\be
N^{{\rm (Max)}}_1=N^{{\rm (Max)}}_2=\cdots=N^{{\rm (Max)}}_k=\frac{N}{k}.
\ee
Its size grows as $\frac{N!}{(N/k)!^k} \sim k^{N}$ up to subexponential factors. Blocks for sectors with non uniform occupation numbers are exponentially smaller. Therefore, if one first takes the large $N$ limit at the level of leading exponential accuracy, only the sector with approximately uniform occupation numbers survives.

This gives a useful way to understand the apparent one dimensionality found at the leading order. The operations of taking the large $N$ limit and imposing the gauge  $S_{N}$ constraint do not commute if the large $N$ limit is taken only at leading exponential order. If we first take the  large $N$ limit in the ungauged Gram matrix, the largest block dominates exponentially over all other blocks. After imposing the gauge constraint, this dominant block gives only a single gauge invariant state, so the resulting Gram matrix appears effectively one dimensional,
\be
G^{{\rm Gauged}} \sim {\rm diag}(1,0,0,\cdots),
\ee
at leading exponential order. 
Since we identify the sector with the uniform occupation numbers $\hm[\vec{p}_{\rm Max}]$ with the largest macroscopic closed universe in the bulk, this prescription amounts to first focusing on the largest macroscopic closed universe sector, then imposing the gauge constraint. 
Thus, again we conclude that the bulk Hilbert space of this closed universe contains only one state.

On the other hand, if we impose the gauge constraint at finite $N$ first, each occupation sector gives one independent state. The Gram matrix is then diagonal, and given by \eqref{eq;GaugedGram}.
Of course the correct prescription consistent with AdS/CFT is this second one.

Thus, we conclude that each (approximate) superselection sector contains only one state.  This is precisely the conclusion obtained by the formal gravitational path integral calculation including the contributions of wormholes. In particular \cite{Harlow:2025pvj} argued that  for any positive integer $n$, 
\be
\overline{\Big({\rm tr}\; (G)^{n} - ({\rm tr} G )^{n} \Big)^{2}} = \sum_{\alpha} p_{\alpha} \Big({\rm tr}\; (G_{\alpha})^{n} - ({\rm tr} G_{\alpha} )^{n} \Big)^{2}=0,
\ee
where averaging is taken over the super selection sector with the index $\alpha$. This implies that for each sector we have rank 1 Gram matrix $G_{\alpha}$, which precisely agrees with our CFT calculation \eqref{eq;GaugedGram}.

In summary, the apparent one dimensionality of the Hilbert space arises only if one first discards all polynomially many subleading sectors in the large $N$ expansion. If these sectors are kept, the gauge invariant Hilbert space remains small but nontrivial.

\section{Density matrices in the presence of observer degrees of freedom} \label{subsec:External}

We have discussed each $\hm_{\vec{\mathcal{N}}_{\alpha}}$ interpreted as the holographic dual of closed universes in the bulk. However, such a Hilbert space was not gauge invariant, therefore does not appear in the final gauge invariant subspace.

Here we explain the way to make it physical,  by coupling it with an external system $\hm_{{\rm obs}}$, which is identified with the Hilbert space of an observer for the CFT degrees of freedom. Let  $\{|\Psi_{(\vec{\mathcal{N}}_{\alpha},i_{\alpha})} \ra \},i_{\alpha}=1, \cdots ,s(\vN_{\alpha})$ be a set of  basis states of $\hm_{\vec{\mathcal{N}}_{\alpha}}$ defined in \eqref{eq;defsector}. Each of $|\Psi_{(\vec{\mathcal{N}}_{\alpha},i_{\alpha})} \ra $ is not gauge invariant, since  the unitary operator $U(\pi)$ defined in \eqref{eq;action} for the permutation $\pi \in S_{N}$ acts non trivially. To make a gauge invariant state, we multiply it with a state $|\vec{\mathcal{N}}_{\alpha},i_{\alpha} \ra$ in the observer system which is transformed  $U^{-1}(\pi) $ under the permutation  $\pi$.
Moreover we demand 
\be
\la \vec{\mathcal{N}}_{\alpha}, i_{\alpha}|\vec{\mathcal{N}}_{\beta}, j_{\beta}\ra= \delta_{\alpha,\beta} \;\delta_{i_{\alpha},j_{\alpha} }.
\ee

The resulting state $|\tilde{\Psi}_{(\vec{\mathcal{N}}_{\alpha},i_{\alpha})} \ra \equiv|\Psi_{(\vec{\mathcal{N}}_{\alpha},i_{\alpha})} \ra \otimes |\vec{\mathcal{N}}_{\alpha},i_{\alpha} \ra $ is in the bipertite system $\hm_{{\rm Ungauged}} \otimes \hm_{{\rm obs}}$ with $\hm_{{\rm Ungauged}} = \tilde{\hm}^{\otimes N} $.

In this setup, let us pick up the following unnormalized maximally entangled state on the bipartite system 
\be
|\Psi \ra = \sum_{\alpha} \sum_{i_{\al}=1}^{s(\vN_{\alpha})} |\Psi_{(\vec{\mathcal{N}}_{\alpha},i_{\alpha})} \ra \otimes |\vec{\mathcal{N}}_{\alpha}, i_{\alpha} \ra.
\ee

The unnormalized density matrix on the observer system is given by 
\be
\rho= \sum_{\al,\beta} \sum_{i_{\al}=1}^{s(\vN_{\alpha})} \sum_{j_{\beta}=1}^{s(\vN_{\beta})} \la \Psi_{(\vec{\mathcal{N}}_{\beta},j_{\beta})} 
|\Psi_{(\vec{\mathcal{N}}_{\alpha},i_{\alpha})} \ra \;|\vec{\mathcal{N}}_{\al},i_{\al}\ra \la \vec{\mathcal{N}}_{\beta},j_{\beta}|.
\label{eq:obse-den}
\ee
We can simply evaluate its entanglement entropy under the averaging\footnote{One can also evaluate the entanglement entropy and its R\'enyi entropy \textit{without taking average} exactly. Since their evaluations are more complicated than the averaged ones, and their results have the almost same scaling dependence as the averages ones, which are enough to discuss underlying structures of our model in this paper.  Therefore, we do not discuss them in this paper, and we will investigate them in a future paper.} by using the result in subsection \ref{subsec:eff-rank}.
Indeed, one can readily check that the following equality holds
\begin{equation}
    \frac{\overline{\tr\left[ \rho^{n} \right]}}{\left( \overline{\tr[\rho]} \right)^{n}}=\frac{\overline{\tr(G^n)}}{(\tr\overline{G})^n}=\left( \mathrm{rank}_{\mathrm{eff}}^{(n)}(G) \right)^{1-n}.
\end{equation}

Then, the R\'enyi entropy is given by
\begin{equation}
  \begin{aligned}
      \overline{S_n(\rho)} &\coloneqq \frac{1}{1-n} \ln \left( \frac{\overline{\tr\left[ \rho^{n} \right]}}{\left( \overline{\tr[\rho]} \right)^{n}} \right)\\
      &= \frac{1}{1-n} \ln \left( \frac{\overline{\tr(G^n)}}{(\tr\overline{G})^n} \right) = \ln C_{N,k} = \ln \binom{N+k-1}{N} \approx \ln \frac{N^{k-1}}{(k-1)!}, \label{eq;renyi}
  \end{aligned}
\end{equation}
where we used \eqref{eq:eff-rank}. Since this R\'enyi entropy does not depend on the R\'enyi index $n$, the entanglement entropy is also given by the above expression.

Thus, the entropy implies that the observer can see the physically relevant log polynomial scale of independent sectors in the large $N$ limit. 

This structure is closely analogous to the calculation of the entanglement entropy of a state defined on two dimensional de Sitter space coupled to a non gravitating bath \cite{Balasubramanian:2020xqf}. In that case, the Rényi entropies are computed by including the fully connected replica wormhole gluing $n$ different copies of Euclidean de Sitter space. Since connecting \(n\) copies of the Euclidean sphere by the fully connected wormhole does not change the Euler characteristic of the resulting manifold, the Rényi entropy becomes independent of the Rényi index \(n\).

\section{Bulk boundary dictionary and closed universes}
\label{sec:bulk}

Our CFT analysis implies in a theory of quantum gravity on closed universes, one should be careful in trusting the result obtained by the saddle point approximation to the gravitational path integral, if we identify the largest sub Hilbert space in the CFT side with the largest macroscopic universe in the dual gravity description. 
That is to say, the exponential dominance of the macroscopic sector in the ungauged description does not translate into an exponential dominance in the physical Hilbert space.

For example, suppose we want to compute the trace of some operator $\phi$.
In the gravitational description, we fix the boundary conditions for the path integral and sum over all possible saddles consistent with them. In the dominant-saddle approximation, however, only the largest closed-universe saddle is retained,
\begin{equation}
{\rm tr}_{\hm_{\rm gravity}}[\phi] =\f{\int \mathcal{D}g_{\mu\nu}\;  e^{-S_{\rm gravity}[g_{\mu\nu}]} \phi}{\int \mathcal{D}g_{\mu\nu}\;  e^{-S_{\rm gravity}} } \sim \f{\la {\rm HH}| \phi|{\rm HH}  \ra}{\la {\rm HH}| {\rm HH}\ra}, \label{eq;grside}
\end{equation}
where $|{\rm HH}\ra$ is the unnormalized Hartle-Hawking state defined by the gravitational path integral with the boundary condition.
 On the other hand, if we compute the corresponding CFT quantity ${\rm tr}_{{\hm_{{\rm CFT}}}}\mathcal{O}$, related by the AdS/CFT dictionary $\phi \leftrightarrow \mathcal{O}$, the states in other superselection sectors labelled by $\al$ can contribute at the same order, because after imposing the gauge constraint each sector is one dimensional,  
\be
{\rm tr}_{{\hm_{{\rm CFT}}}}\left[\mathcal{O}\right]= \f{\la {\rm HH}| \mathcal{O}|{\rm HH}  \ra}{\la {\rm HH}| {\rm HH}\ra} + \sum_{\al \neq {\rm HH}} \f{\la \vec{\mathcal{N}}_{\al} | \mathcal{O}|\vec{\mathcal{N}}_{\al} \ra}{\la \vec{\mathcal{N}}_{\al} | \vec{\mathcal{N}}_{\al} \ra}, \label{eq;CFTside}
 \ee
 where $|\vec{\mathcal{N}}_{\al} \ra$ is the single state in the super selection sector $\al$. We nevertheless argue that when the CFT degrees of freedom is coupled with external degrees of freedom.

If we naively apply the bulk to boundary dictionary, the two quantities appear to be equal. However, by comparing \eqref{eq;grside} and \eqref{eq;CFTside}
that this cannot be correct. This stems from the following fact. Although the dimension of the Hartle-Hawking sector overwhelms those of the other baby-universe sectors before imposing the gauge constraint, after imposing the gauge constraint each sector becomes one dimensional. Therefore all sectors contribute with the same order of magnitude. In particular, we cannot approximate the answer by the contribution of the Hartle-Hawking sector alone.

Still, we can reproduce the semiclassical answer if we consider the bipartite system
$\hm_{{\rm Extended}}=
\hm_{{\rm Ungauged}}\otimes \hm_{{\rm obs}}.
$
On this Hilbert space, we get
\be
{\rm tr}_{\hm_{{\rm Extended}}}[ \mathcal{O}]
\sim
\frac{\langle {\rm HH}|\phi|{\rm HH}\rangle}{\langle {\rm HH}|{\rm HH}\rangle} = {\rm tr}_{\hm_{\rm bulk}}[\phi].
\ee

\section{Conclusion}\label{sec:conclusion}

We have studied a class of Hilbert spaces in symmetric orbifold CFTs and shown that imposing the $S_N$ gauge constraint drastically changes the structure of the would be closed universe Hilbert space. At leading exponential order in the large $N$ expansion, each closed universe sector appears  one dimensional. If polynomially many subleading sectors are kept, however, the gauge invariant Hilbert space contains $
\binom{N+k-1}{k-1}\sim N^{k-1}
$ independent superselection sectors.
 
 A natural next step is to identify the precise bulk interpretation of this construction. In particular, it would be important to understand what operation in the bulk corresponds to quotienting by the permutation group $S_N$ on the CFT side. In related constructions of de Sitter Hilbert spaces, such as those based on perturbative bulk excitations, one starts with perturbative degrees of freedom described by de Sitter QFT and then imposes the gravitational constraints by gauging the de Sitter isometry group. Our construction has a similar logic, but the relevant gauge group is instead the permutation group acting on the copies of the seed theory. It would therefore be very interesting to identify the bulk symmetry, or redundancy, corresponding to this permutation gauging.

One possible clue comes from the role of wormholes in gravitational path integrals. In many gravitational calculations involving several replicated systems, such as computations with multiple black holes or replica geometries, the gravitational path integral includes wormhole saddles connecting different replicas~\cite{Penington:2019kki,Almheiri:2019qdq,Marolf:2020xie,Saad:2019lba}. Including these wormhole contributions often has the effect of enforcing an additional gauge equivalence among semiclassical configurations. From this perspective, it is tempting to speculate that the $S_N$ gauging in the symmetric orbifold CFT is related, on the bulk side, to a wormhole induced gauging or redundancy among different universe sectors.

Clarifying this relation would also shed light on the nature of the superselection sectors that survive in the bulk. In our CFT construction, these sectors are labeled by occupation number data and remain distinguishable after imposing the gauge constraint. Understanding their bulk counterparts may provide a microscopic interpretation of the superselection structure of closed universes, and may help explain how the apparently one dimensional Hilbert space obtained from leading semiclassical gravitational path integrals is refined by subleading, genuinely quantum degrees of freedom.

\subsection*{Almost completely evaporated black holes and BH/String correspondence}
Our analysis of closed universes is also closely related to the physics of black holes near the end of evaporation. In such a regime, the black hole originally had an entropy of order $e^{1/G_N}$ states, or equivalently an entropy of order \(1/G_N\), but the remaining number of degrees of freedom becomes much smaller, potentially only polynomial in \(1/G_N\). This is precisely the kind of regime in which the leading semiclassical gravitational description is expected to become insufficient, while subleading quantum degrees of freedom may still survive.

 Symmetric orbifold CFTs are known to be closely related to tensionless string theory in AdS$_3$. This suggests that the closed universe sectors studied in this paper may admit a natural interpretation in string theoretic terms.

Combining these observations, it is tempting to speculate that black holes near the endpoint of evaporation may be naturally described in terms of stringy degrees of freedom. In this sense, our symmetric orbifold construction may provide a useful microscopic arena for studying the transition from a semiclassical black hole description to a string theoretic description.

It would be particularly interesting to make this connection more precise. Doing so may provide a microscopic AdS/CFT realization of the correspondence between black holes and highly excited strings~\cite{Susskind:1993ws,Horowitz:1996nw}. From this perspective, the polynomially many closed universe sectors found in our analysis may capture a regime in which the black hole Hilbert space has already lost its semiclassical exponential entropy, but still retains a nontrivial stringy structure.

\subsection*{Introducing observers and Entwinement}

The closed universe sectors constructed in the first part of this paper are not gauge invariant by themselves. We argued that they can nevertheless be made physical by coupling them to observer degrees of freedom, so that the combined system satisfies the gauge constraint. A natural way to implement this idea in the present setup is to divide the original ungauged system into two parts: among the $N$ copies of the seed Hilbert space, one may identify $M$ copies with observer degrees of freedom and the remaining $N-M$ copies with the closed universe degrees of freedom. The physical Hilbert space is then obtained by imposing the full $S_N$ gauge constraint on the combined system.

This viewpoint is closely related to recent discussions of observers in closed universes. In particular, it has been argued that semiclassical physics for an observer can be recovered with an accuracy controlled by the entanglement entropy between the observer and the closed universe degrees of freedom. Thus, in our setup, it is important to understand how much entanglement can exist between the observer sector and the closed universe sector after imposing the permutation gauge constraint.

In symmetric orbifold theories, entanglement associated with gauge degrees of freedom has been discussed under the name of ``Entwinement''~\cite{Balasubramanian:2014sra,Balasubramanian:2018ajb}. It would therefore be interesting to clarify the role of entwinement in our construction. In particular, understanding how entwinement between different subsets of the ($S_N$) gauged degrees of freedom controls the emergence of observer dependent semiclassical physics may provide a useful bridge between the CFT construction and the bulk description of closed universes.

\section*{Acknowledgements}
We thank Takato Mori and Kotaro Tamaoka for early collaboration, Masamichi Miyaji, Yasunori Nomura, and Ronak Soni for discussions. This work was supported in part by MEXT KAKENHI Grant-in-Aid for Transformative Research Areas A “Extreme Universe” No. 21H05184.
H.T. was supported by the "circulation program for young researchers" in this Grant-in-Aid.
A.M. was also supported by JSPS KAKENHI Grant Number JP26KJ0186. T.U was also supported in part by MEXT KAKENHI Grant Number JP25K00997 and 25K07289. H.T was also supported by the “THERS Make New Standard Program for the Next Generation Researchers”. Part of this work was completed during the workshop "Holographic Universe 2026" at Yukawa Institute of Theoretical Physics (YITP). We thank the institute for its hospitality and the organizers and participants for stimulating discussions.

\appendix

\section{Permutation averages and auxiliary Gaussian derivations}
\label{app:permutation-averages}

In this appendix, we provide (alternative) derivations of several results in the main text. Some of them were originally obtained by combinatorial arguments.

\subsection{Quantities and results}

Let $N$ and $k$ be positive integers with $N>k$ as considered in the main body of this paper, and let $S_N$ be the permutation group of degree $N$. We consider the product
\begin{equation*}
  \delta_{x_1,x_{\pi(1)}}\delta_{x_2,x_{\pi(2)}}\cdots\delta_{x_N,x_{\pi(N)}},
\end{equation*}
where $x_1,\ldots,x_N \in \{1,\ldots,k\}, \pi \in S_{N}$. For any function $F(x_1,\ldots,x_N;\pi)$, we denote the uniform average over the indices, with $\pi$ kept fixed, by
\begin{equation}
  \overline{F(x_1,\ldots,x_N;\pi)} \coloneqq \frac{1}{k^N} \sum_{x_1,\ldots,x_N=1}^{k}F(x_1,\ldots,x_N;\pi).
\end{equation}

The first quantity of interest is
\begin{equation}\label{eq:Asf_def}
  \begin{aligned} \mathsf{A}_{N,k} &\coloneqq \frac{1}{|S_N|} \sum_{\pi\in S_N} \overline{\delta_{x_1,x_{\pi(1)}}\delta_{x_2,x_{\pi(2)}}\cdots\delta_{x_N,x_{\pi(N)}}} \\
  &= \frac{1}{k^N|S_N|} \sum_{\pi\in S_N}\sum_{x_1,\ldots,x_N=1}^{k} \delta_{x_1,x_{\pi(1)}}\delta_{x_2,x_{\pi(2)}}\cdots\delta_{x_N,x_{\pi(N)}}. \end{aligned}
\end{equation}
The result derived below is
\begin{equation}\label{eq:Asf_res}
  \mathsf{A}_{N,k}=\frac{1}{k^N}\binom{N+k-1}{N}.
\end{equation}

We next consider two independent index configurations $\vec{x}=(x_1,\ldots,x_N)$ and $\vec{y}=(y_1,\ldots,y_N)$. Define
\begin{equation}\label{eq:Bsf_def}
  \begin{aligned} \mathsf{B}_{N,k} &\coloneqq \frac{1}{|S_N|} \sum_{\pi\in S_N} \overline{\delta_{x_1,y_{\pi(1)}}\delta_{x_2,y_{\pi(2)}}\cdots\delta_{x_N,y_{\pi(N)}}} \\
  &= \frac{1}{k^{2N}|S_N|} \sum_{\pi\in S_N} \sum_{x_1,\ldots,x_N=1}^{k} \sum_{y_1,\ldots,y_N=1}^{k} \delta_{x_1,y_{\pi(1)}}\delta_{x_2,y_{\pi(2)}}\cdots\delta_{x_N,y_{\pi(N)}}. \end{aligned}
\end{equation}
The result is
\begin{equation}\label{eq:Bsf_res}
  \mathsf{B}_{N,k}=\frac{1}{k^N}.
\end{equation}

We also consider the schematic two-copy quantity
\begin{equation}\label{eq:Csf_def}
  \begin{aligned} \mathsf{C}_{N,k} &\coloneqq \frac{1}{|S_N|^2} \sum_{\pi,\tau\in S_N} \overline{\left( \delta_{x_1,y_{\pi(1)}}\cdots\delta_{x_N,y_{\pi(N)}} \right) \left( \delta_{y_1,x_{\tau(1)}}\cdots\delta_{y_N,x_{\tau(N)}} \right)} \\
  &= \frac{1}{k^{2N}|S_N|^2} \sum_{\pi,\tau\in S_N} \sum_{x_1,\ldots,x_N=1}^{k} \sum_{y_1,\ldots,y_N=1}^{k} \left( \delta_{x_1,y_{\pi(1)}}\cdots\delta_{x_N,y_{\pi(N)}} \right) \left( \delta_{y_1,x_{\tau(1)}}\cdots\delta_{y_N,x_{\tau(N)}} \right). \end{aligned}
\end{equation}
It becomes
\begin{equation}\label{eq:Csf_res}
  \mathsf{C}_{N,k}=\frac{1}{k^{2N}}\binom{N+k-1}{N}.
\end{equation}

Next, to analyze fluctuations of the permutation sum itself, we introduce
\begin{equation}\label{eq:Dsf_def}
  \begin{aligned} \mathsf{D}_{N,k} &\coloneqq \overline{ \left( \frac{1}{|S_N|} \sum_{\pi\in S_N} \delta_{x_1,x_{\pi(1)}}\cdots\delta_{x_N,x_{\pi(N)}} \right)^2 } \\
  &= \frac{1}{k^N|S_N|^2} \sum_{x_1,\dots,x_N=1}^{k} \sum_{\pi_1,\pi_2\in S_N} \prod_{a=1}^{N} \delta_{x_a,x_{\pi_1(a)}}\delta_{x_a,x_{\pi_2(a)}}. \end{aligned}
\end{equation}
The generating-functional derivation below gives
\begin{equation}\label{eq:Dsf_res}
  \mathsf{D}_{N,k} = \frac{1}{k^N N!} \sum_{n_1+\cdots+n_k=N} n_1!\cdots n_k!.
\end{equation}

Finally, we will use an $n$-copy cyclic generalization of the independent two-copy overlap $\mathsf{C}_{N,k}$. Introduce $n$ independent index configurations
\begin{equation*}
  \vec{x}^{(\alpha)} = (x^{(\alpha)}_1,\ldots,x^{(\alpha)}_N), \qquad \alpha=1,\ldots,n,
\end{equation*}
under the cyclic identification $\vec{x}^{(n+1)}\equiv \vec{x}^{(1)}$. We first define the unnormalized cyclic contraction sum over all possible configurations and permutations:
\begin{equation}\label{eq:Esf_unnormalized_def}
  \mathsf{E}_{N,k}^{(n)} \coloneqq \sum_{\vec{x}^{(1)},\ldots,\vec{x}^{(n)}=1}^k \prod_{\alpha=1}^{n} \left[ \sum_{\pi_\alpha\in S_N} \prod_{a=1}^{N} \delta_{x^{(\alpha)}_a, x^{(\alpha+1)}_{\pi_\alpha(a)}} \right].
\end{equation}
The exact sequential Gaussian integration derived below yields
\begin{equation}\label{eq:Esf_unnormalized_res}
  \mathsf{E}_{N,k}^{(n)} = (N!)^n\binom{N+k-1}{N}.
\end{equation}

We also define $\mathsf{F}_{N,k}^{(n)}$ as the fully normalized joint ensemble average. This quantity can be naturally expressed in two equivalent ways: first, as the macroscopic ensemble average denoted by the overline, and second, as the microscopic full summation over the independent unconstrained configurations and the permutation operations. Explicitly, we define
\begin{equation}\label{eq:Fsf_def}
  \begin{aligned} \mathsf{F}_{N,k}^{(n)} &\coloneqq \overline{ \frac{1}{|S_N|^n} \sum_{\pi_1, \dots, \pi_n \in S_N} \prod_{\alpha=1}^{n} \prod_{a=1}^{N} \delta_{x^{(\alpha)}_a, x^{(\alpha+1)}_{\pi_\alpha(a)}} } \\
  &= \frac{1}{k^{nN}} \sum_{\vec{x}^{(1)},\ldots,\vec{x}^{(n)}=1}^k \left\{ \prod_{\alpha=1}^n \left( \frac{1}{|S_N|} \sum_{\pi_\alpha\in S_N} \right)  \prod_{a=1}^{N} \delta_{x^{(\alpha)}_a, x^{(\alpha+1)}_{\pi_\alpha(a)}} \right\} \\
  &= \frac{1}{k^{nN}(N!)^n} \mathsf{E}_{N,k}^{(n)}, \end{aligned}
\end{equation}
where the overline in the first line represents the uniform index average over the $k^{nN}$ states. Because the complete summation over all multi-indices eliminates any remaining free index, the structural overline maps directly and exactly onto the layer of nested sums. By substituting the exact evaluation of $\mathsf{E}_{N,k}^{(n)}$ into this joint average, the permutation group volumes completely cancel out, leaving the exact normalized result:
\begin{equation}\label{eq:Fsf_res}
  \mathsf{F}_{N,k}^{(n)} = \frac{1}{k^{nN}}\binom{N+k-1}{N}.
\end{equation}
For $n=2$, under the boundary identification $\vec{x}^{(1)}\equiv \vec{x}$ and $\vec{x}^{(2)}\equiv \vec{y}$, this cyclic chain correctly recovers the independent two-copy overlap probability $\mathsf{C}_{N,k}$ in \eqref{eq:Csf_res}.

\subsection{Derivation of \texorpdfstring{$\mathsf{A}_{N,k}$}{A(N,k)}}

First, introduce $k$ free complex scalar variables $\vec{\phi}=(\phi_1,\ldots,\phi_k),\, \bar{\vec{\phi}}=(\bar\phi_1,\ldots,\bar\phi_k),$ and the normalized Gaussian measure
\begin{equation}
  d\mu(\phi,\bar\phi) \coloneqq \prod_{a=1}^{k}\frac{d\phi_a d\bar\phi_a}{\pi} \exp\left(-\sum_{a=1}^{k}|\phi_a|^2\right).
\end{equation}
The Gaussian average is
\begin{equation}
  \langle F(\vec{\phi},\bar{\vec{\phi}})\rangle \coloneqq \int d\mu(\phi,\bar\phi)\,F(\vec{\phi},\bar{\vec{\phi}}).
\end{equation}
The basic Wick contraction reads
\begin{equation*}
  \langle \phi_x\bar\phi_y\rangle=\delta_{xy},
\end{equation*}
and more generally,
\begin{equation*}
  \begin{aligned} \langle \phi_{x_1}\phi_{x_2}\bar\phi_{y_1}\bar\phi_{y_2}\rangle &= \delta_{x_1y_1}\delta_{x_2y_2}+\delta_{x_1y_2}\delta_{x_2y_1} = \sum_{\sigma\in S_2}\delta_{x_1,y_{\sigma(1)}}\delta_{x_2,y_{\sigma(2)}}. \end{aligned}
\end{equation*}
For $N$ pairs, Wick's theorem gives
\begin{equation}\label{eq:Wick_N}
  \langle \phi_{x_1}\cdots\phi_{x_N}\bar\phi_{y_1}\cdots\bar\phi_{y_N}\rangle = \sum_{\pi\in S_N}\delta_{x_1,y_{\pi(1)}}\cdots\delta_{x_N,y_{\pi(N)}}.
\end{equation}
Setting $y_m=x_m$ therefore gives
\begin{equation}\label{eq:Wick_N_diag}
  \sum_{\pi\in S_N} \delta_{x_1,x_{\pi(1)}}\cdots\delta_{x_N,x_{\pi(N)}} = \left\langle \phi_{x_1}\cdots\phi_{x_N}\bar\phi_{x_1}\cdots\bar\phi_{x_N} \right\rangle.
\end{equation}
Substituting this identity into the definition \eqref{eq:Asf_def} yields
\begin{equation}
  \begin{aligned} \mathsf{A}_{N,k} &= \frac{1}{k^N|S_N|} \sum_{x_1,\ldots,x_N=1}^{k} \left\langle \phi_{x_1}\cdots\phi_{x_N}\bar\phi_{x_1}\cdots\bar\phi_{x_N} \right\rangle \\
  &= \frac{1}{k^N|S_N|} \left\langle \left(\sum_{a=1}^{k}|\phi_a|^2\right)^{N} \right\rangle. \end{aligned}
\end{equation}
To evaluate the remaining integral, introduce the radial variable
\begin{equation*}
  r^2=\sum_{a=1}^{k}|\phi_a|^2.
\end{equation*}
Since the volume for angular directions is
\begin{equation*}
  \mathrm{Vol}(S^{2k-1})=\frac{2\pi^k}{(k-1)!},
\end{equation*}
we obtain
\begin{equation*}
  \begin{aligned} \left\langle \left(\sum_{a=1}^{k}|\phi_a|^2\right)^{N} \right\rangle &= \frac{1}{\pi^k}\int_{\mathbb{C}^k}d^{2k}\phi\,e^{-r^2}r^{2N} \\
  &= \frac{1}{\pi^k}\frac{2\pi^k}{(k-1)!} \int_{0}^{\infty}dr\,r^{2k-1}e^{-r^2}r^{2N}. \end{aligned}
\end{equation*}
With the change of variables $t=r^2$, $dt=2rdr$,
\begin{equation*}
  \int_{0}^{\infty}dr\,r^{2(k+N)-1}e^{-r^2} = \frac{1}{2}\int_{0}^{\infty}dt\,t^{k+N-1}e^{-t} =\frac{1}{2}(k+N-1)!.
\end{equation*}
Hence,
\begin{equation}\label{eq:phi_moment}
  \left\langle \left(\sum_{a=1}^{k}|\phi_a|^2\right)^{N} \right\rangle = \frac{(k+N-1)!}{(k-1)!}.
\end{equation}
Since $|S_N|=N!$, substituting \eqref{eq:phi_moment} back gives
\begin{equation}
  \mathsf{A}_{N,k} = \frac{1}{k^N N!}\frac{(k+N-1)!}{(k-1)!} = \frac{1}{k^N}\binom{N+k-1}{N}.
\end{equation}

\subsection{Derivation of \texorpdfstring{$\mathsf{B}_{N,k}$}{B(N,k)}}

We evaluate $\mathsf{B}_{N,k}$. Using the contraction identity \eqref{eq:Wick_N} already obtained in the derivation of $\mathsf{A}_{N,k}$,
\begin{equation*}
  \sum_{\pi\in S_N} \delta_{x_1,y_{\pi(1)}}\cdots\delta_{x_N,y_{\pi(N)}} = \left\langle \phi_{x_1}\cdots\phi_{x_N}\bar\phi_{y_1}\cdots\bar\phi_{y_N} \right\rangle,
\end{equation*}
we obtain
\begin{equation}
  \begin{aligned} \mathsf{B}_{N,k} &= \frac{1}{k^{2N}|S_N|} \sum_{x_1,\ldots,x_N=1}^{k} \sum_{y_1,\ldots,y_N=1}^{k} \left\langle \phi_{x_1}\cdots\phi_{x_N}\bar\phi_{y_1}\cdots\bar\phi_{y_N} \right\rangle \\
  &= \frac{1}{k^{2N}|S_N|} \left\langle \left(\sum_{x=1}^{k}\phi_x\right)^N \left(\sum_{y=1}^{k}\bar\phi_y\right)^N \right\rangle. \end{aligned}
\end{equation}
Now set
\begin{equation*}
  X\coloneqq \sum_{x=1}^{k}\phi_x.
\end{equation*}
To evaluate its moment, it is convenient to make a unitary change of variables in $\mathbb{C}^k$. Choose a unitary matrix $U\in U(k)$ whose first row is
\begin{equation*}
  \frac{1}{\sqrt{k}}(1,1,\ldots,1),
\end{equation*}
and define new variables
\begin{equation*}
  \psi_a\coloneqq \sum_{x=1}^{k}U_{ax}\phi_x, \qquad a=1,\ldots,k.
\end{equation*}
Since the transformation is unitary, the Jacobian is $1$ and
\begin{equation*}
  \sum_{x=1}^{k}|\phi_x|^2=\sum_{a=1}^{k}|\psi_a|^2.
\end{equation*}
Moreover, by construction,
\begin{equation*}
  \psi_1=\frac{1}{\sqrt{k}}\sum_{x=1}^{k}\phi_x=\frac{X}{\sqrt{k}}, \qquad\text{so} \qquad X=\sqrt{k}\,\psi_1.
\end{equation*}
Therefore
\begin{equation}
  \begin{aligned} \left\langle |X|^{2N}\right\rangle &= k^N \int \prod_{a=1}^{k}\frac{d\psi_a d\bar\psi_a}{\pi} \exp\left(-\sum_{a=1}^{k}|\psi_a|^2\right)|\psi_1|^{2N} \\
  &= k^N \left( \int\frac{d\psi_1 d\bar\psi_1}{\pi}e^{-|\psi_1|^2}|\psi_1|^{2N} \right) \prod_{a=2}^{k} \left( \int\frac{d\psi_a d\bar\psi_a}{\pi}e^{-|\psi_a|^2} \right). \end{aligned}
\end{equation}
The integrals over $\psi_2,\ldots,\psi_k$ are equal to $1$, so only the first mode remains. Writing $\psi_1=re^{i\theta}$, so that $d\psi_1 d\bar\psi_1=r\,dr\,d\theta$, and then setting $t=r^2$, we obtain
\begin{equation*}
  \begin{aligned} \left\langle |X|^{2N}\right\rangle &= k^N\frac{1}{\pi}\int_{0}^{2\pi}d\theta\int_{0}^{\infty}dr\,e^{-r^2}r^{2N+1} \\
  &= 2k^N\int_{0}^{\infty}dr\,e^{-r^2}r^{2N+1} = k^N\int_{0}^{\infty}dt\,e^{-t}t^N \\
  &= k^N N!. \end{aligned}
\end{equation*}
Therefore,
\begin{equation}
  \mathsf{B}_{N,k} = \frac{1}{k^{2N}N!}\,k^N N! = \frac{1}{k^N}.
\end{equation}

\subsection{Derivation of \texorpdfstring{$\mathsf{C}_{N,k}$}{C(N,k)}}

Introduce two independent copies of the complex Gaussian variables,
\begin{equation*}
  \vec{\phi}^{(1)}=(\phi_1^{(1)},\ldots,\phi_k^{(1)}), \qquad \vec{\phi}^{(2)}=(\phi_1^{(2)},\ldots,\phi_k^{(2)}),
\end{equation*}
and abbreviate
\begin{equation*}
  \vec{\phi}\coloneqq \vec{\phi}^{(1)}, \qquad \vec{\varphi}\coloneqq \vec{\phi}^{(2)}.
\end{equation*}
Starting from the Kronecker-delta expression in \eqref{eq:Csf_def}, we apply \eqref{eq:Wick_N} separately to the two permutation sums. This gives
\begin{equation*}
  \sum_{\pi\in S_N} \delta_{x_1,y_{\pi(1)}}\cdots\delta_{x_N,y_{\pi(N)}} = \left\langle \phi_{x_1}^{(1)}\cdots\phi_{x_N}^{(1)} \bar\phi_{y_1}^{(1)}\cdots\bar\phi_{y_N}^{(1)} \right\rangle_1,
\end{equation*}
and
\begin{equation*}
  \sum_{\tau\in S_N} \delta_{y_1,x_{\tau(1)}}\cdots\delta_{y_N,x_{\tau(N)}} = \left\langle \phi_{y_1}^{(2)}\cdots\phi_{y_N}^{(2)} \bar\phi_{x_1}^{(2)}\cdots\bar\phi_{x_N}^{(2)} \right\rangle_2.
\end{equation*}
Because the two Gaussian copies are independent, the product of these two averages is the joint Gaussian average of the product. Therefore
\begin{equation*}
  \sum_{\pi,\tau\in S_N} \prod_{a=1}^{N} \delta_{x_a,y_{\pi(a)}}\delta_{y_a,x_{\tau(a)}}= \left\langle \prod_{a=1}^{N} \phi_{x_a}^{(1)}\bar\phi_{y_a}^{(1)} \phi_{y_a}^{(2)}\bar\phi_{x_a}^{(2)} \right\rangle.
\end{equation*}
Substituting this into \eqref{eq:Csf_def} and then carrying out the sums over the independent indices $x_a$ and $y_a$ gives
\begin{equation}\label{eq:Csf_gaussian_two_copy}
  \begin{aligned} \mathsf{C}_{N,k} &= \frac{1}{k^{2N}|S_N|^2} \sum_{x_1,\ldots,x_N=1}^{k} \sum_{y_1,\ldots,y_N=1}^{k} \left\langle \prod_{a=1}^{N} \phi_{x_a}^{(1)}\bar\phi_{y_a}^{(1)} \phi_{y_a}^{(2)}\bar\phi_{x_a}^{(2)} \right\rangle \\
  &= \frac{1}{|S_N|^2}\frac{1}{k^{2N}} \left\langle \left(\sum_{x=1}^{k}\phi_x^{(1)}\bar\phi_x^{(2)}\right)^N \left(\sum_{y=1}^{k}\phi_y^{(2)}\bar\phi_y^{(1)}\right)^N \right\rangle. \end{aligned}
\end{equation}
Using
\begin{equation*}
  \sum_{x=1}^{k}\phi_x^{(1)}\bar\phi_x^{(2)}=\vec{\phi}\cdot\bar{\vec{\varphi}}, \qquad \sum_{y=1}^{k}\phi_y^{(2)}\bar\phi_y^{(1)}=\vec{\varphi}\cdot\bar{\vec{\phi}},
\end{equation*}
we need to evaluate
\begin{equation}
  \left\langle \left(\vec{\phi}\cdot\bar{\vec{\varphi}}\right)^N \left(\vec{\varphi}\cdot\bar{\vec{\phi}}\right)^N \right\rangle.
\end{equation}
For this purpose, it is useful to prove a slightly more general Gaussian identity. Let $\vec{A}$ and $\vec{B}$ be fixed $k$-component vectors. Then
\begin{equation}\label{eq:Gaussian_AB_identity}
  \left\langle \left(\vec{A}\cdot\vec{\varphi}\right)^N \left(\bar{\vec{\varphi}}\cdot\vec{B}\right)^N \right\rangle_{\varphi} = N!\,(\vec{A}\cdot\vec{B})^N.
\end{equation}
Indeed, introduce the generating function
\begin{equation}
  Z(\alpha,\beta) = \left\langle \exp\left( \alpha\,\vec{A}\cdot\vec{\varphi}+\beta\,\bar{\vec{\varphi}}\cdot\vec{B} \right) \right\rangle_{\varphi}.
\end{equation}
Using the normalized complex Gaussian measure, we complete the square as
\begin{equation}
  -\bar{\vec{\varphi}}\cdot\vec{\varphi}+\alpha\,\vec{A}\cdot\vec{\varphi}+\beta\,\bar{\vec{\varphi}}\cdot\vec{B} = -(\bar{\vec{\varphi}}-\alpha\vec{A})\cdot(\vec{\varphi}-\beta\vec{B})+\alpha\beta\,\vec{A}\cdot\vec{B}.
\end{equation}
Therefore,
\begin{equation}
  Z(\alpha,\beta)=\exp\left(\alpha\beta\,\vec{A}\cdot\vec{B}\right).
\end{equation}
Taking $N$ derivatives with respect to both $\alpha$ and $\beta$ at $\alpha=\beta=0$, we obtain
\begin{equation}
  \begin{aligned} \left\langle \left(\vec{A}\cdot\vec{\varphi}\right)^N \left(\bar{\vec{\varphi}}\cdot\vec{B}\right)^N \right\rangle_{\varphi} &= \left.\frac{\partial^N}{\partial\alpha^N}\frac{\partial^N}{\partial\beta^N}Z(\alpha,\beta)\right|_{\alpha=\beta=0} \\
  &= N!\,(\vec{A}\cdot\vec{B})^N. \end{aligned}
\end{equation}
We now apply this identity with
\begin{equation*}
  \vec{A}=\bar{\vec{\phi}}, \qquad \vec{B}=\vec{\phi},
\end{equation*}
and integrate over $\vec{\varphi}$ first. This gives
\begin{equation}
  \left\langle \left(\bar{\vec{\phi}}\cdot\vec{\varphi}\right)^N \left(\bar{\vec{\varphi}}\cdot\vec{\phi}\right)^N \right\rangle_{\varphi} = N!\,\left(\bar{\vec{\phi}}\cdot\vec{\phi}\right)^N = N!\,|\vec{\phi}|^{2N}.
\end{equation}
This is the same as the desired expression, up to the order of the scalar products. Therefore the full two-copy Gaussian average becomes
\begin{equation}
  \left\langle \left(\vec{\phi}\cdot\bar{\vec{\varphi}}\right)^N \left(\vec{\varphi}\cdot\bar{\vec{\phi}}\right)^N \right\rangle = N!\left\langle |\vec{\phi}|^{2N}\right\rangle.
\end{equation}
Using \eqref{eq:phi_moment},
\begin{equation}
  \left\langle |\vec{\phi}|^{2N}\right\rangle = \frac{(N+k-1)!}{(k-1)!},
\end{equation}
we obtain
\begin{equation}
  \left\langle \left(\vec{\phi}\cdot\bar{\vec{\varphi}}\right)^N \left(\vec{\varphi}\cdot\bar{\vec{\phi}}\right)^N \right\rangle = N!\frac{(N+k-1)!}{(k-1)!}.
\end{equation}
Substituting back into \eqref{eq:Csf_gaussian_two_copy}, we find
\begin{equation}
  \begin{aligned} \mathsf{C}_{N,k} &= \frac{1}{(N!)^2k^{2N}}\,N!\frac{(N+k-1)!}{(k-1)!} \\
  &= \frac{1}{k^{2N}}\binom{N+k-1}{N}. \end{aligned}
\end{equation}
Thus, the two-copy computation gives the same binomial factor as \eqref{eq:Asf_res}, with the extra factor $k^{-N}$ coming from the second independent index average in \eqref{eq:Csf_def}.

\subsection{Derivation of the cyclic \texorpdfstring{$n$}{n}-copy contraction}

We now derive \eqref{eq:Esf_unnormalized_res}. Introduce $n$ independent copies of complex Gaussian variables
\begin{equation*}
  \vec{\phi}^{(\alpha)} = (\phi^{(\alpha)}_1,\ldots,\phi^{(\alpha)}_k), \qquad \alpha=1,\ldots,n.
\end{equation*}
Applying Wick's theorem separately to each permutation sum gives
\begin{equation}
  \mathsf{E}_{N,k}^{(n)} = \left\langle \prod_{\alpha=1}^{n} \left( \sum_{x=1}^{k} \bar{\phi}^{(\alpha)}_x \phi^{(\alpha+1)}_x \right)^N \right\rangle ,
\end{equation}
where $\vec{\phi}^{(n+1)}\equiv\vec{\phi}^{(1)}$.
This expression is the cyclic $n$-copy generalization of \eqref{eq:Csf_gaussian_two_copy}: the two-copy contraction is replaced by nearest-neighbor contractions between consecutive copies, with the last copy contracted back to the first.

We now use the Gaussian identity \eqref{eq:Gaussian_AB_identity} iteratively. First integrate the Gaussian variable $\vec{\phi}^{(2)}$. The factors involving $\vec{\phi}^{(2)}$ are
\begin{equation*}
  \left( \bar{\vec{\phi}}^{(1)}\cdot\vec{\phi}^{(2)} \right)^N \left( \bar{\vec{\phi}}^{(2)}\cdot\vec{\phi}^{(3)} \right)^N .
\end{equation*}
Using \eqref{eq:Gaussian_AB_identity}, this integration gives
\begin{equation*}
  N! \left( \bar{\vec{\phi}}^{(1)}\cdot\vec{\phi}^{(3)} \right)^N .
\end{equation*}
Repeating this for $\vec{\phi}^{(3)},\ldots,\vec{\phi}^{(n)}$ gives
\begin{equation}
  \mathsf{E}_{N,k}^{(n)} = (N!)^{n-1} \left\langle \left( \bar{\vec{\phi}}^{(1)}\cdot\vec{\phi}^{(1)} \right)^N \right\rangle .
\end{equation}
The remaining integral is
\begin{equation}
  \left\langle \left( \bar{\vec{\phi}}^{(1)}\cdot\vec{\phi}^{(1)} \right)^N \right\rangle = \frac{(N+k-1)!}{(k-1)!} = N!\binom{N+k-1}{N}.
\end{equation}
Therefore,
\begin{equation}
  \mathsf{E}_{N,k}^{(n)} = (N!)^n\binom{N+k-1}{N}.
\end{equation}
Dividing by $k^{nN}(N!)^n$ gives
\begin{equation}\label{eq:FNk}
  \mathsf{F}_{N,k}^{(n)} = \frac{1}{k^{nN}}\binom{N+k-1}{N}.
\end{equation}

\subsection{Derivation of \texorpdfstring{$\mathsf{D}_{N,k}$}{D(N,k)}}

Using the Wick-theorem correspondence \eqref{eq:Wick_N_diag}, the product of two delta chains can be represented by two independent Gaussian copies as
\begin{equation*}
  \sum_{\pi_1,\pi_2\in S_N} \prod_{a=1}^{N}\delta_{x_a,x_{\pi_1(a)}}\delta_{x_a,x_{\pi_2(a)}} = \left\langle \phi_{x_1}^{(1)}\bar{\phi}_{x_1}^{(1)}\cdots\phi_{x_N}^{(1)}\bar{\phi}_{x_N}^{(1)} \phi_{x_1}^{(2)}\bar{\phi}_{x_1}^{(2)}\cdots\phi_{x_N}^{(2)}\bar{\phi}_{x_N}^{(2)} \right\rangle.
\end{equation*}
After summing over $x_1,\dots,x_N$, the quantity of interest \eqref{eq:Dsf_def} becomes
\begin{equation*}
  \mathsf{D}_{N,k} \coloneqq \frac{1}{k^N|S_N|^2} \left\langle \left(\sum_{x=1}^{k}|\phi_x^{(1)}|^2|\phi_x^{(2)}|^2\right)^N \right\rangle.
\end{equation*}
To evaluate this moment systematically, introduce the generating functional
\begin{equation}\label{eq:Z_def}
  Z(\alpha) \coloneqq \left\langle \exp\left(\alpha\sum_{x=1}^{k}|\phi_x^{(1)}|^2|\phi_x^{(2)}|^2\right) \right\rangle.
\end{equation}
Because the Gaussian measure factorizes over each mode $m=1,\dots,k$, we have
\begin{equation*}
  Z(\alpha) = \prod_{m=1}^{k}\left\{ \int\frac{d\phi_m^{(1)}d\bar{\phi}_m^{(1)}}{\pi} \frac{d\phi_m^{(2)}d\bar{\phi}_m^{(2)}}{\pi} e^{- |\phi_m^{(1)}|^2 - |\phi_m^{(2)}|^2 + \alpha |\phi_m^{(1)}|^2 |\phi_m^{(2)}|^2} \right\}.
\end{equation*}
Writing $\phi_m^{(a)}=r_a e^{i\theta_a}$, the angular integrations are trivial and cancel the factors of $\pi$ in the normalized measure. With $t_a=r_a^2$, the single-mode integral becomes
\begin{equation*}
  \begin{aligned} \int_0^\infty dt_1\,e^{-t_1}\int_0^\infty dt_2\,e^{-t_2(1-\alpha t_1)} &= \int_0^\infty dt_1\,e^{-t_1}\frac{1}{1-\alpha t_1} \\
  &= \int_0^\infty dt_1\,e^{-t_1}\sum_{n=0}^\infty (\alpha t_1)^n \\
  &= \sum_{n=0}^\infty n!\,\alpha^n. \end{aligned}
\end{equation*}
Hence,
\begin{equation}\label{eq:Z_res}
  Z(\alpha) = \prod_{m=1}^{k}\left(\sum_{n=0}^\infty n!\,\alpha^n\right) = \sum_{n_1,\dots,n_k=0}^\infty n_1!\cdots n_k!\,\alpha^{n_1+\cdots+n_k}.
\end{equation}
The $N$-th moment is obtained from the $N$-th derivative of \eqref{eq:Z_res} at $\alpha=0$:
\begin{equation*}
  \begin{aligned} \left\langle \left(\sum_{x=1}^{k}|\phi_x^{(1)}|^2|\phi_x^{(2)}|^2\right)^N \right\rangle &= \left.\frac{d^N}{d\alpha^N}Z(\alpha)\right|_{\alpha=0} \\
  &= N!\sum_{n_1+\cdots+n_k=N} n_1!\cdots n_k!. \end{aligned}
\end{equation*}
Substituting this into the definition of $\mathsf{D}_{N,k}$ gives
\begin{equation}
  \mathsf{D}_{N,k} = \frac{1}{k^N(N!)^2} \cdot N!\sum_{n_1+\cdots+n_k=N}n_1!\cdots n_k! = \frac{1}{k^N N!} \sum_{n_1+\cdots+n_k=N}n_1!\cdots n_k!.
\end{equation}

\section{Calculations  of the variance}\label{app:variance}

In section \ref{sec:one-dim}, we computed the averaged Gram matrix $\;\overline{\langle \Psi_{\vec{x}} 
| \Psi_{\vec{y}} \rangle}$ of the projected states. 
Now we compute its variance.  First, focusing on the diagonal part (the norm), its expression involves two permutations $\pi, \sigma \in S_{N}$, 
\be
\overline{\;\langle \Psi_{\vec{x}} 
| \Psi_{\vec{x}} \rangle^{2}} = \frac{1}{\mathcal{M}^{2}}\cdot \f{1}{k^{N}|S_{N}|^{2}} \sum_{\pi, \sigma \in S_{N}} \sum_{x_{1}, \cdots ,x_{N}=1}^{k} \prod_{a=1}^{N} 
\delta_{x_{a},x_{\pi(a)}}
\delta_{x_{a},x_{\sigma(a)}}.
\ee

The sum over $\pi, \sigma \in S_{N}$ can again be computed by passing over to the occupation number representation\footnote{One can also use the result in appendix \ref{app:permutation-averages} to obtain the result.}, 
\be
\sum_{\pi, \sigma \in S_{N}} 
\prod_{a=1}^{N} 
\delta_{x_{a},x_{\pi(a)}}
\delta_{x_{a},x_{\sigma(a)}} =\left(\prod_{a=1}^{k} N_{a}! \right)^{2}.
\ee
Thus,
\be
\overline{\;\langle \Psi_{\vec{x}} 
| \Psi_{\vec{x}} \rangle^{2}} = \frac{1}{\mathcal{M}^{2}}\cdot\f{1}{k^{N}|S_{N}|}\sum_{\substack{N_{1}, \cdots N_{k} \\ N_{1}+\cdots N_{k}=N} } \left(\prod_{a=1}^{k} N_{a} !\right).
\ee
In the large $N$ limit, the right hand side is dominated by configurations in which a single occupation number equals to $N$, and all the remaining occupation numbers are zero. Therefore, we obtain
\be
\overline{\;\langle \Psi_{\vec{x}} 
| \Psi_{\vec{x}} \rangle^{2}}  \approx  \frac{1}{\mathcal{M}^{2}}\cdot \f{1}{k^{N-1}} \approx \frac{ \left( (k-1)! \right)^{2} \; k^{N+1} }{N^{2k-2}},
\ee
where, in the second approximation, we used the large $N$ limit of the normalization factor \eqref{eq:normalization-factor-M}. Then, the variance of the norm is given by
\begin{equation}
\f{\overline{\;\left(\langle \Psi_{\vec{x}} 
| \Psi_{\vec{x}} \rangle -\overline{\;\langle \Psi_{\vec{x}} 
| \Psi_{\vec{x}} \rangle} 
\right)^{2}}}{\left(\overline{\;\langle \Psi_{\vec{x}} 
| \Psi_{\vec{x}} \rangle} \right)^{2}} \approx \frac{ \left( (k-1)! \right)^{2} \; k^{N+1} }{N^{2k-2}}-1.
\end{equation}

We can similarly compute the off-diagonal part (the overlap),
\begin{align}
\overline{\;\langle \Psi_{\vec{x}} 
| \Psi_{\vec{y}} \rangle^{2}} &=\frac{1}{\mathcal{M}^{2}}\cdot \f{1}{k^{N}|S_{N}|^{2}} \sum_{\pi, \sigma \in S_{N}} \sum_{x_{1}, \cdots ,x_{N}=1}^{k} \sum_{y_{1}, \cdots ,y_{N}=1}^{k}\prod_{a=1}^{N} 
\delta_{x_{a},y_{\pi(a)}}
\delta_{x_{a},y_{\sigma(a)}} \\[+10pt]
&=\frac{1}{\mathcal{M}^{2}}\cdot\f{1}{k^{2N}} \begin{pmatrix}
 N+k-1 \\ N
\end{pmatrix}=\frac{1}{C_{N,k}}, 
\end{align}
where we used \eqref{eq:normalization-factor-M}, \eqref{eq:Csf_def} and \eqref{eq:Csf_res}. Here,  $C_{N,k}$ is the binomial coefficient \eqref{eq:bi-nomial}.
Therefore, the normalized variance is given by 
\be
\f{\overline{\;\left(\langle \Psi_{\vec{x}} 
| \Psi_{\vec{y}} \rangle -\overline{\;\langle \Psi_{\vec{x}} 
| \Psi_{\vec{y}} \rangle} 
\right)^{2}}}{\left(\overline{\;\langle \Psi_{\vec{x}} 
| \Psi_{\vec{y}} \rangle} \right)^{2}}= \dfrac{\dfrac{1}{C_{N,k}} \left[1-\dfrac{1}{C_{N,k}} \right]}{ \left(\dfrac{1}{C_{N,k}}\right)^{2} }=C_{N,k}-1.
\ee

\bibliographystyle{JHEP}
\bibliography{deSitter}

\end{document}